\documentclass[preprint,review,12pt,authoryear]{elsarticle}

\usepackage{amssymb}
\usepackage{amsmath}
\usepackage{graphicx}
\usepackage{subfig}
\usepackage{epstopdf}
\usepackage[top=25mm,bottom=25mm,left=25mm,right=20mm,headsep=10mm]{geometry}
\usepackage{indentfirst}
\usepackage{bm}

\journal{}

\begin{document}

\begin{frontmatter}



\title{A Continuum Multi-Disconnection-Mode Model \\
for Grain Boundary Migration}


\author[a,b]{Chaozhen Wei}
\ead{iasczwei@ust.hk}
\author[c]{Spencer L. Thomas}
\author[c]{Jian Han}
\author[c,d,e]{David J. Srolovitz}
\ead{srol@cityu.edu.hk}
\author[a]{Yang Xiang}
\ead{maxiang@ust.hk}

\address[a]{Department of Mathematics, The Hong Kong University of Science and Technology, Clear Water Bay, Kowloon, Hong Kong SAR, China}
\address[b]{Institute for Advanced Study, The Hong Kong University of Science and Technology, Clear Water Bay, Kowloon, Hong Kong SAR, China}
\address[c]{Department of Materials Science and Engineering, University of Pennsylvania, Philadelphia, PA 19104-6272 USA}
\address[d]{Department of Mechanical Engineering and Applied Mechanics, University of Pennsylvania, Philadelphia, PA 19104-6272 USA}
\address[e]{Department of Materials Science and Engineering, City University of Hong Kong, Kowloon, Hong Kong SAR, China}


\begin{abstract}
We study the Grain Boundary (GB) migration based on the underlying disconnection structure and mechanism. Disconnections are line defects that lie solely within a GB and are characterized by both a Burgers vector and a step height, as set by the GB bicrystallography. Multiple disconnection modes can nucleate, as determined by their formation energy barriers and temperature, and move along the GB under different kinds of competing driving forces including shear stress and chemical
potential jumps across the GBs. We present a continuum model in two dimensions for GB migration where the GB migrates via the thermally-activated nucleation and kinetically-driven motion of disconnections. We perform continuum numerical simulations for investigating the GB migration behavior in single and multi-mode disconnection limits in both a bicrystal (under two types of boundary conditions) and for a finite-length GB with pinned ends. The results clearly demonstrate the significance of including the coupling and competing between different disconnection modes and driving forces for describing the complex and diverse phenomena of GB migration within polycyrstalline microstructures.
\end{abstract}


\end{frontmatter}


\section{INTRODUCTION}
\label{sec:intro}
The microstructure of a polycrystalline material can be abstracted as a network of grain boundaries (GBs) --  interfaces between differently oriented crystalline grains.
Hence, GB migration is synonymous  with polycrystalline microstructure evolution.
Since the evolution of such microstructures strongly affect many mechanical, thermal and electronic properties of polycrystalline materials \citep{sutton1995interfaces}, understanding  GB migration mechanisms, dynamics and kinetics is essential to tailoring the structure and properties of polycrystalline materials.

In  conventional  capillarity-driven grain growth theory \citep{Mullins1956two,hillert1965on},  grain boundary migration is described as motion by mean curvature $v_{\bot}=M_{gb}\gamma \kappa$ (for the isotropic case), where $M_{gb}$ is the temperature-dependent GB mobility, $\gamma$ is the GB energy density (surface tension) and $\kappa$ is the GB mean curvature.
Grain boundaries migrate in the direction of their local normal such as to reduce the local (and total) GB energy (GB area).
In the isotropic case, GBs meet at triple junctions (TJs) with a dihedral angle $2\pi/3$ (assuming they are in or near equilibrium) leading to the famous von Neumann-Mullins relation for grain size evolution in two-dimensions \citep{von1952metal,Mullins1956two} and the MacPherson-Srolovitz formula in three-dimensions \citep{macpherson2007neumann}.
There is an extensive literature on applying capillarity-driven GB motion to simulate grain growth using continuum methods (e.g., see \citep{chenlq1994simulation,Kinderlehrer2006variation,elsey2009diffusion,lazar2010accurate}) and to analyze the results of  atomistic simulations \citep{upmanyu1998}.
While the isotropic, capillarity-driven grain boundary migration theory has been widely applied, both GB mobility and GB energy depend on the misorientation between grains and the inclination of the grain boundary plane \citep{Read1950} (in three dimensions this is a five-dimensional parameter space).
The effects of such anisotropy on GB migration and  grain growth has also been widely examined \citep{kazaryan2000phase,upmanyu2002simulation,zhang2005curvature} as has its effects on  abnormal grain growth \citep{rollett1989abnormal,decost1017abnormal} and grain rotation during migration \citep{harris1998rotation,Kobayashi2000continuum,upmanyu2006simulation,esedoglu2016grain}.
However, this generalized capillarity-driven model fails to explain many widely observed GB migration phenomena associated affected by mechanical stresses; for example, stress-driven grain growth \citep{legros2008abnormal,JIN2004direct}, grain boundary sliding \citep{VanSwygenhoven2002dislocationemission,legros2008abnormal,SCHAFER2012}, grain rotation \citep{JIN2004direct,Ma2004Science,Shan2004rotation} and abnormal grain growth \citep{Simpson1971,Riontino1979}.

While the effects of stress on GB migration have long been recognized  \citep{li1953stress,Bainbridge1954shear},  it is only relatively recently that such stress-coupled GB migration has been recognized as a general phenomenon  \citep{srinivasan2002rotation,cahn2006coupling}.
In shear-coupled GB migration, the motion of the GB in the direction of its normal (i.e., migration) is coupled to a tangential translation of one grain with respect to the other meeting at the GB.
Cahn and Taylor \citep{cahn2004unified,taylor2007accommodation} proposed a description that couples curvature-driven GB motion to mechanical stresses and describes  grain boundary sliding and grain rotation.
\cite{zhang2018lowangle} developed a continuum model for shear coupling in low-angle GBs in terms of the motion and reaction of the constituent dislocations that constitute the GB structure.

Substantial experimental and atomistic simulation evidence exists for the presence of shear-coupled migration in high-angle GBs \citep{winning2001stressinduced,gottstein2001grain,winning2002mechanism,rupert2009science,Molteni1996prl,Molteni1997prl,Hamilton2002prb,chenlq1992MD,Shiga2004MD,Sansoz2005MD,Trautt2012MD,Homer2013MD}.
\cite{zhang2018lowangle} developed a continuum model for shear coupling in low angle GBs in terms of the motion and reaction of the constituent dislocations that constitute the GB structure.
Recent experiments and simulations demonstrated that the shear-coupling of high-angle GBs  \citep{Rajabzadeh2013evidence,rajabzadeh2013elementary,Mompiou2015coupling} is associated with the motion of line defects known as disconnections \citep{bollmann1970crystal,Ashby1972498,hirth1996disconnection}.
Disconnections are constrained to lie within the GB and are characterized by both a Burgers vector $\bm{b}$ and a step height $H$.
The motion of disconnections along the GB leads to both GB migration (associated with the step height) and shear coupling (associated with the Burgers vectors)  \citep{thomas2017reconciling,han2018}.
Recently, a continuum model for GB migration was proposed based upon the motion of a single type of disconnection \citep{zhang2017prl}.
However, bicrystallography allows for an infinite, discrete set of possible disconnection types or modes \citep{king1980,han2018}.
A disconnection mode is associated with a Burgers vector and step height pair $\{\bm{b}_n,H_{nj}\}$; for each Burgers vector $\bm{b}_n$ there are an infinite set of possible step heights $\{H_{nj}\}$.
The selection of and competition between disconnection modes are central to understanding  the temperature-dependence of GB shear coupling, mobility, and sliding   \citep{thomas2017reconciling,han2018,chen2019shear}.
Such competition varies between GBs and even along a GB within a microstructure, resulting in complex/rich GB migration phenomenon  in polycrystalline microstructures.

In the present paper, we  propose a continuum formulation for  GB dynamics that accounts for the complexity/richness associated with multiple disconnection modes and apply it by performing continuum simulations that elucidate the interplay between disconnection mode selection, mechanical boundary conditions, driving forces and temperature during GB migration.
In order to validate the predictions, we compare these predictions with atomistic (molecular dynamics, MD) simulation results \citep{thomas2017reconciling}.
This paper is organized as follows.
In Section~\ref{sec:model}, we present a continuum model for GB migration based upon multiple disconnection modes.
Next, we discuss the treatment of the resulting elasticity problem in a bicrystal with prescribed boundary conditions.
In Section~\ref{sec:simulations}, we perform simulations of shear-coupled GB migration in a bicrystal with two different types of boundary conditions as a function of temperature and in a finite-length GB (e.g., a GB  constrained by  triple junctions).
Finally, in Section~\ref{sec:summary}, we discuss the implication of these results and identify some outstanding questions in GB dynamics in real microstructures.

\section{MATHEMATICAL MODEL}
\label{sec:model}
\subsection{Grain Boundary Motion with Multiple Disconnection Modes}
\label{subsec:GB_model}
\begin{figure}[ht]
\centering
\includegraphics[width=0.75\textwidth]{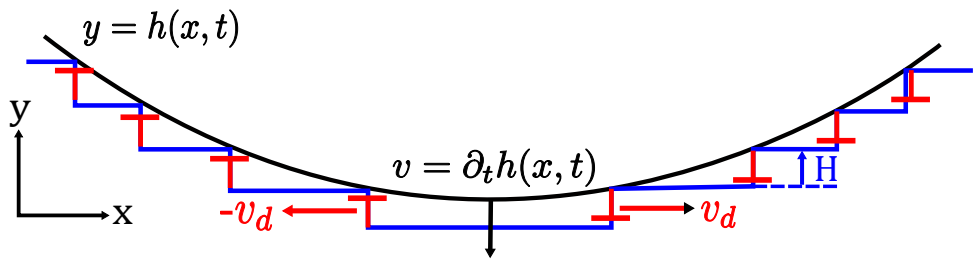}
\caption{A schematic illustration of shear-coupled migration of a general (small slope) curved GB via the motion of disconnections.
The normal migration of the GB, with velocity $v$ (in the $y$-direction), is achieved by disconnection  glide motion  with velocity $v_d$ (in the $x$-direction). }\label{fig:GB_scheme}
\end{figure}

Consider a GB of general shape $y=h(x,t)$, as shown in Fig.~\ref{fig:GB_scheme}.
Assuming that the GB profile deviates only slightly from that of a reference GB (i.e., a flat GB with its normal parallel to the $y$-axis),  $|\partial_x h|\ll1$, we may describe the GB profile in terms of an  array of disconnections with Burgers vector along the $x$-direction and step height along the $y$-direction.
(Note, the GB, in its reference configuration has its own structure as described by a combination of structural units with a small period along the $x$-axis \citep{sutton1995interfaces,han2017redux}.)
In the disconnection mechanism of GB migration, the GB migration rate is controlled by the motion of disconnections along the GB (we focus on disconnection glide here and do not explicitly consider disconnection climb, although there are circumstances for which it is important).
When only one disconnection mode is operating, a positive disconnection $(\bm{b},H)$ gliding in the $+x$ direction (or a negative disconnection $(-\bm{b},-H)$ gliding in the $-x$ direction) will move the GB segment in its wake down by $H$ and shift the top grain with respect to the bottom grain by $b$ (the Burgers vector parallel to the $x$-direction).
In this way, GB migration is coupled with the lateral shear translation of one grain with respect to the other meeting at the GB; this is disconnection-mediated, shear-coupled GB migration.

A continuum model for GB migration via the glide  of a single disconnection mode was previously proposed \citep{zhang2017prl}.
Here, we  generalize the description of GB motion to account for multiple disconnection modes (including the competition between these modes).
Suppose that the GB ($y=h(x,t)$) in Fig.~\ref{fig:GB_scheme} contains multiple disconnection modes $(\bm{b}_n, H_{nj})$ with $\bm{b}_n=(b_n,0)$ parallel to the $x$-axis and $H_{nj}$ parallel to the $y$-axis, where $n$ is the Burgers vector index and $j$ represents one of the allowed step heights corresponding to $\bm{b}_n$.
For simplicity, and without loss of generality, we write the set of possible disconnection modes as  $\{\bm{b}_i, H_{i}\}$ where  the subscript $i$ is the disconnection mode index.
The disconnection density $\rho_i$ is positive/negative to represent the density of disconnection of type $(\bm{b}_i, H_i)$/$(-\bm{b}_i, -H_i)$.
GB migration may now be described in terms of the disconnection flux $J_i$
\begin{eqnarray}
\label{eq:migration}
\frac{\partial h}{\partial t} = -\sum_i J_iH_i.
\end{eqnarray}
A continuity condition insures conservation of Burgers vector and step height
\begin{eqnarray}
\label{eq:evolution}
\frac{\partial \rho_i}{\partial t} + \frac{\partial J_i}{\partial x}=0.
\end{eqnarray}

The disconnection flux $J_i$ is defined as
\begin{eqnarray} \label{eq:flux}
J_i = v_i (|\rho_i| + 2c_i),
\end{eqnarray}
where $v_i$ is the glide velocity of a positive disconnection and $c_i$ is the thermal equilibrium concentration of disconnections of type $i$.
The latter term represents a model for disconnection nucleation \citep{zhang2017prl}
\begin{equation}
c_i = \frac{1}{a} e^{-E^*_i/(k_B T)},
\label{eq:nucleation}
\end{equation}
where $a$ is an atomic spacing, $k_B$ is the Boltzmann constant, $T$ is temperature and $E^*_i$ is half of the formation energy of a disconnection pair \citep{han2018}.
We assume that disconnection nucleation is sufficiently facile that the nucleation rate is determined by thermal equilibrium considerations (this may not  be universally valid).
The relation between the GB shape and disconnection density is simply $\partial_x h = \sum_i \rho_i H_i$ (consistent with Eqs.~(\ref{eq:migration})-(\ref{eq:evolution})).
During  shear-coupled GB migration, disconnection glide induces a shear translation between the two grains meeting at the GB.
The relative grain translation (GB sliding) rate $\partial_t B$ and the shear coupling factor are related to the disconnection flux
\begin{eqnarray}
\label{eq:translation}
&\partial_t B& = \sum_i J_ib_i, \\
&\beta& = \partial_t B/\partial_t h,
\end{eqnarray}
where $\beta$ is the shear coupling factor defined as the ratio of the grain translation rate and the GB migration rate and is an  important descriptor of  shear-coupled GB migration.

We assume that disconnection motion is overdamped, such that disconnection velocity is  proportional to the driving force.
For a disconnection of mode ($\bm{b},H$), $v_d = M_d\cdot f_d$, where $M_d$ is the disconnection mobility and $f_d$ is the total driving force on this disconnection.
The driving force consists of two parts $f_{d}=f_{b}+f_h$, where $f_b$ is the force conjugate to the Burgers vector and $f_h$ to the step character.
The driving force associated on the disconnection associated with a stress is  the Peach-Koehler force \citep{PK_force1950} along the glide direction, i.e., $f_b=[(\bm{\sigma}\cdot\bm{b})\times\bm{\xi}]\cdot\bm{g}$, where $\bm{\sigma}$ is the local stress (tensor), $\bm{\xi}$ is the disconnection line direction (perpendicular to the $x$-$y$ plane) and $\bm{g}$ is the glide direction (i.e.,  $x$).
The driving force associated with the step character is related to the jump in chemical potential across the GB, i.e., $f_h=(\Psi-\gamma \partial_{xx} h)H$ where $\Psi$ is the difference between the energy densities in two grains meeting at a GB (e.g., the synthetic driving force widely used in atomistic simulations of GB migration) and where we have explicitly separated out the contribution to the chemical potential jump associated with capillarity ($\gamma$ is the GB energy and $\partial_{xx} h$ is  the GB mean curvature).
These driving forces can be derived directly from the variation of the total energy of the system with respect to the virtual displacement of the disconnection along the GB; i.e.,  $f_b$ and $f_h$ correspond to the variation of the (long-range interaction) elastic energy and the non-elastic contributions to the energy associated with disconnection motion.
Inserting these forces into the expression for the disconnection velocity yields
\begin{eqnarray}\label{eq:v_d}
v_d = M_d [(\sigma+\tau)b + (\Psi-\gamma \partial_{xx} h)H],
\end{eqnarray}
where $\sigma$ and $\tau$ are shear components of the internal and applied stress. Equations~(\ref{eq:evolution}-\ref{eq:v_d}) represent a closed system.

Combining Eqs.~(\ref{eq:migration},\ref{eq:flux},\ref{eq:v_d}) yields an equation of motion for the GB migration with multiple disconnection modes
\begin{eqnarray}
\label{eq:multiEOM}
\frac{\partial h}{\partial t} = -\sum_i M_{i} [(\sigma+\tau)b_i +(\Psi-\gamma \partial_{xx} h)H_i] (|\rho_i| + 2c_i)H_i,
\end{eqnarray}
where $M_i$ is the mobility of a disconnection of mode $i$.
While different disconnection modes will, in general, have different mobilities (with different temperature dependency), for the sake of simplicity of presentation, here we assume  that all disconnections have the same, constant mobility $M_i=M_d$.
(Disconnection mobilities will, in general, be temperature-dependent with activation energies that depend on disconnection type, local bonding, GB structure, solute segregation, point defects, etc.)
Note that the expression for the GB velocity does not explicitly depend on a GB mobility.
In other words, rather than the GB velocity simply being the product of the driving force on the GB and a GB mobility, it is  determined by the properties of the disconnections (i.e., their mobilities  \underline{AND} their Burgers vector and step height) and their densities along the GB.

The relative importance of the different disconnection modes depends on their relative ease of nucleation under the local conditions, as represented by the parameter $c_i$ in Eq.~\eqref{eq:multiEOM} and dependent on both  disconnection formation energy $E^*_i$ and temperature (see Eq.~\eqref{eq:nucleation}).
The formation energy (per unit length) of a disconnection dipole is $2E^*_i = \tilde{A}|H_i|+\tilde{B}|b_i|^2$  \citep{han2018} (for a straight dislocation dipole in a periodic system), where $\tilde{A}$ is the excess energy density due to the step and $\tilde{B}$ includes the disconnection core energy  and  the elastic interaction energy between the two members of the disconnection pair.
The coefficients $\tilde{A}$ and $\tilde{B}$ can be either estimated analytically based upon a continuum model \citep{han2018} or determined by fitting to  atomistic simulation \citep{chen2019shear} or experimental results.
We note that accurate determination of these values should account  for the temperature dependence of the formation energy  \citep{LeSar1989finite,yang2015}.

\begin{figure}[!ht]
\centering
\subfloat[]{\includegraphics[width=0.16\textwidth]{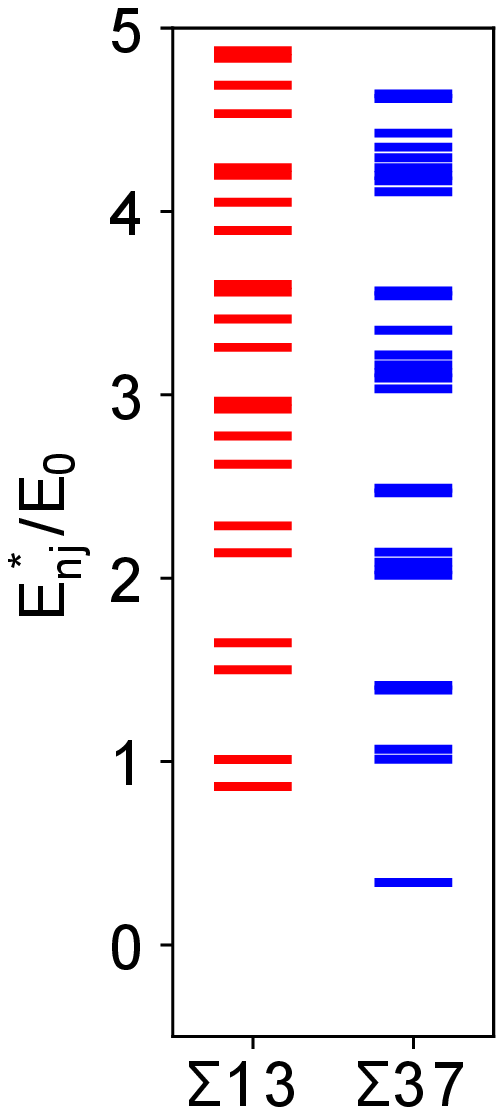}}
\subfloat[]{\includegraphics[width=0.42\textwidth]{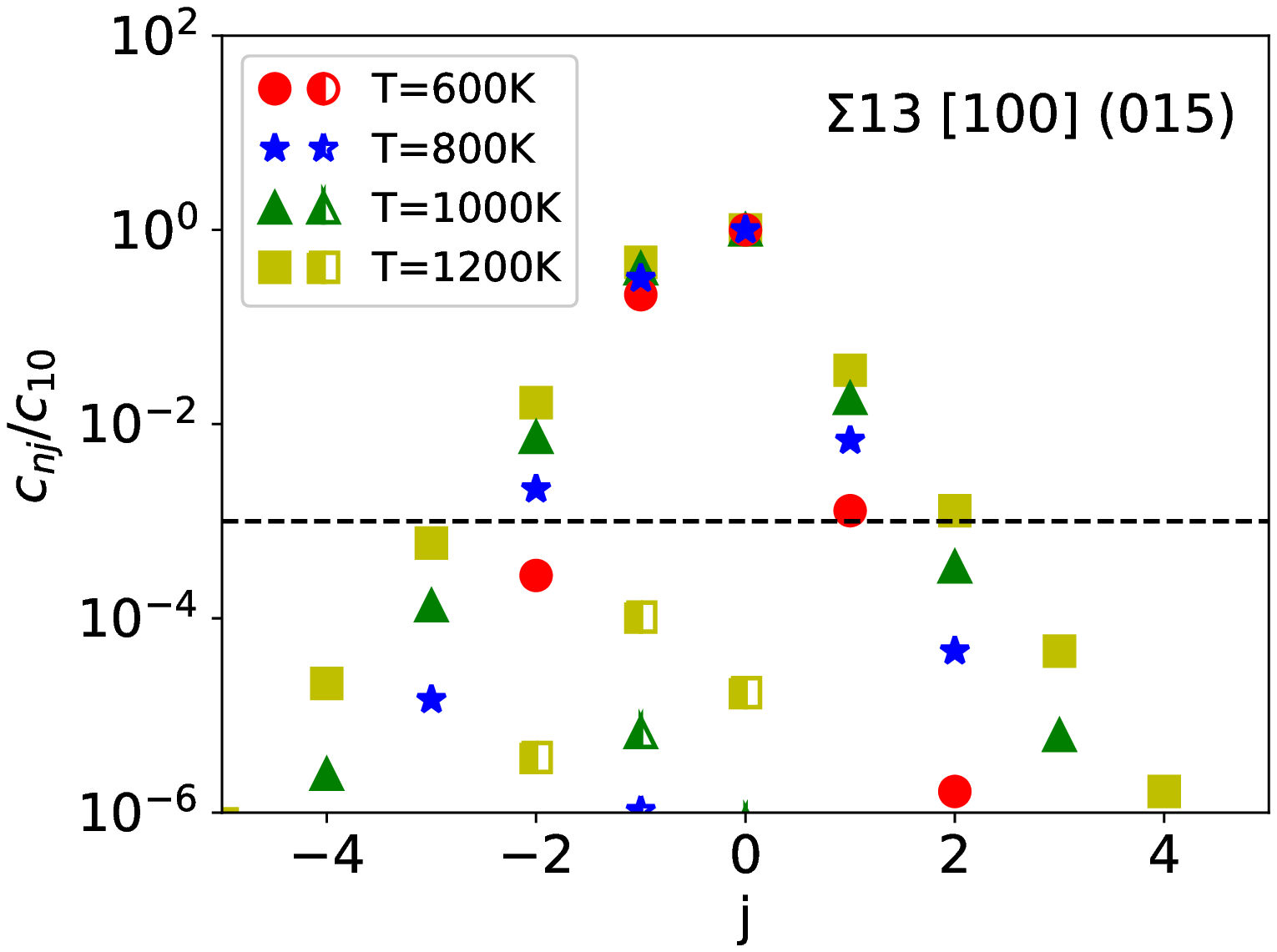}}
\subfloat[]{\includegraphics[width=0.42\textwidth]{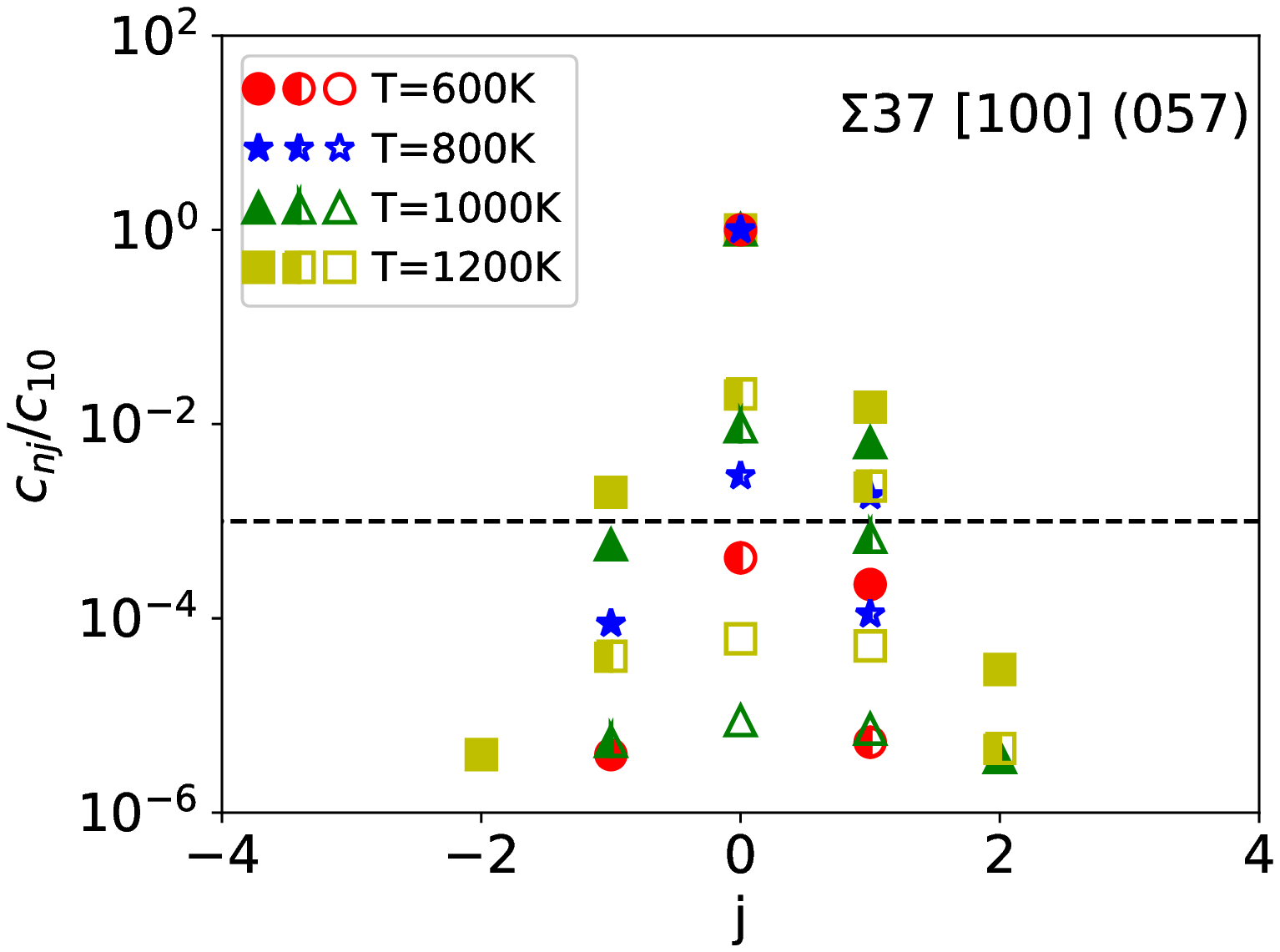}}
\caption{Formation energy and nucleation rates of possible disconnection modes of $\Sigma$13 and $\Sigma$37 STGBs in Cu. (a) The formation energy $E^*_{nj}$ of disconnection modes $(b_n,H_{nj})$ for $n=1,2,\cdots$ and $j=0,\pm1,\pm2,\cdots$, where $E_0=\tilde{A}a_0$ sets the overall energy scale and $a_0$ is the lattice constant. (b) and (c) Nucleation rates $c_{nj}$ of disconnection modes ($b_n$,$H_{nj}$) at $T=600$, 800, 1000, 1200~K for $\Sigma$13 and $\Sigma$37 GBs, respectively. Filled, half-filled and empty symbols represent disconnection modes with $n=1,2,$ and $3$, respectively. The disconnection modes with $n>3$ have smaller nucleation rates and are beyond the range of this figure. The horizontal dashed line ($c_{nj}/c_{10}=10^{-3}$) is an arbitrary scale that is used to indicate the relative importance of different disconnection modes.} \label{fig:dis_energy}
\end{figure}

The simultaneous activation of multiple  disconnection modes adds complexity to GB migration behavior.
Since the disconnection nucleation rate is dependent on both formation energy and temperature,  the ``apparent'' GB mobility will vary between GBs, with temperature, and with the nature of the driving force for GB migration \citep{chen2019shear}.
We explicitly consider the example of $\Sigma13$  and $\Sigma37$ symmetric tilt GBs in Cu.
For these GBs, simple analysis of the crystallography \citep{han2018} shows that for the $\Sigma$13 GB, $b_n=na_0/\sqrt{26}$ and $H_{nj}=-(5n+13j)a_0/2\sqrt{26}$ and for the $\Sigma$37 GB, $b_n=na_0/\sqrt{74}$ and $H_{nj}=(6n-37j)a_0/2\sqrt{74}$.
We extract the values of $\tilde{A}$ and $\tilde{B}$ from \cite{han2018} and \cite{chen2019shear}: $\tilde{A}_{\Sigma13}=0.48$  and $\tilde{A}_{\Sigma37}=0.53$ J/m$^2$ and  $\tilde{B}_{\Sigma13}=42$ and $\tilde{B}_{\Sigma37}=36$ GPa.
Figure~\ref{fig:dis_energy} shows the formation energies $E^*_{nj}$ and nucleation rates $c_{nj}$ for these two GBs (as a function of temperature).

The disconnection mode $(b_1,H_{10})$ has the largest nucleation rate $c_{10}$ (lowest formation energy) and hence makes the primary contribution to the shear-coupled GB migration at low temperature, while other disconnection modes, with much smaller nucleation rates, compared to the primary disconnection mode make much smaller contributions (at least at low temperature).
As shown in Figs.~\ref{fig:dis_energy}(b) and \ref{fig:dis_energy}(c), the disconnection modes  above the horizontal dashed line (arbitrarily set to $c_{nj}/c_{10}>10^{-3}$) are considered to be significant during shear-coupled GB migration; while those disconnection modes below the dashed line ($c_{nj}/c_{10}<10^{-3}$) are considered relatively insignificant here.

For the $\Sigma$37 GB (Fig.~\ref{fig:dis_energy}(c)), at low temperature ($T=600$~K) the primary disconnection mode ($c_{10}$) dominates the shear-coupled GB migration ($c_{nj}/c_{10}<10^{-3}$ for the other modes), and hence the corresponding coupling factor of the GB migration should be  $\beta\simeq-b_{1}/H_{10}$.
At higher temperatures, in addition to the primary disconnection mode, other disconnection modes  become significant (with higher formation energy  $c_{nj}/c_{10}>10^{-3}$) and are expected to play a role during  GB migration.
In contrast,  shear-coupled GB migration of the $\Sigma$13 GB at low temperature ($T=600$~K) will be governed by two disconnection modes, $c_{10}$ and $c_{1\overline{1}}$ (where bars over indices indicates negative values ) as shown in Fig.~\ref{fig:dis_energy}(b), owing to the small gap between the formation energies of these two disconnection modes (see Fig.~\ref{fig:dis_energy}(a)).

When multiple disconnection modes make significant contributions to GB migration, the shear coupling is not determined by a single disconnection mode but rather the average effect of all operating disconnection modes in response to all local driving forces \citep{thomas2017reconciling,chen2019shear}.
In general, the coupling factor $\beta$ for shear-coupled GB migration depends on the mode-specific properties of the disconnection, types of driving forces, and temperature, subject to environment constraints (e.g., mechanical boundary conditions).
The present model for GB migration based upon multiple disconnection modes enables us to describe the diverse shear coupling behavior of GB migration associated with the competition between these disconnection modes.
For example, it can be applied to  interesting high temperature phenomena as pure grain boundary sliding ($\beta\rightarrow \infty$) and pure GB migration without shear deformation ($\beta\rightarrow 0$) which cannot be captured by the single disconnection model for GB migration \citep{zhang2017prl}.

\subsection{Boundary Conditions}
\label{subsec:Elasticity_BCs}
In real polycrystalline microstructures, each grain is confined by surrounding grains and each GB is delimited by junctions with three other GBs (triple junctions, TJs); hence, migration of individual GBs is influenced by other GBs and TJs
Shear displacements that accompany GB migration propagate within the grains, but limited by constraints associated with other grains.
Disconnections cannot, in general, be transmitted through TJ.
These  constraints can result in stress generation and accumulation within grains and/or at TJs during shear-coupled GB migration in the polycrystal.
The complexity of polycrystalline systems hinders the complete analysis of these  constraints on GB migration.
Here, we  probe the effects of the  constraints by investigating the migration of individual GBs in  bicrystals with imposed boundary conditions at the top and bottom surfaces (this provides an analog to the constraints imposed from other grains) or in a finite-length GB with pinned ends (an analog to TJ constraints).
For the former case,  we consider two types of bicrystal boundary conditions.
\begin{figure}[!b]
\centering
\includegraphics[width=0.5\textwidth]{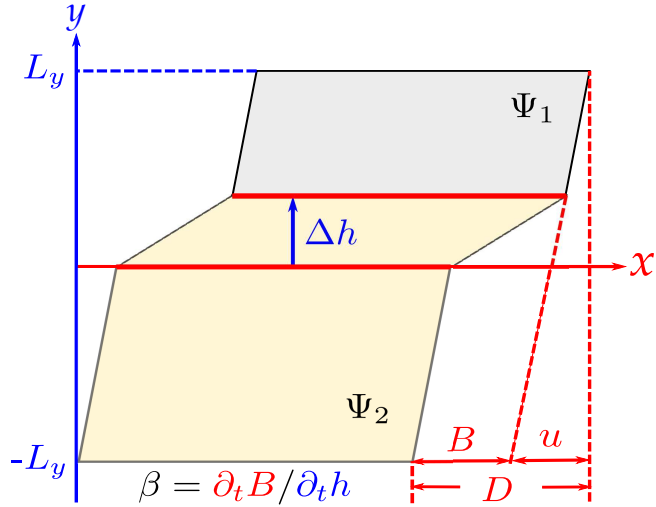}
\caption{A schematic illustration of shear-coupled GB migration within a bicrystal.
When the GB migrates by $\Delta h$, the relative translational displacement between the top ($y=L_y$) and bottom ($y=-L_y$) surfaces is $D=B+u$, where $u$ is the elastic deformation and $B$ is the ``plastic'' deformation (grain translation) induced by the glide  of disconnections along GBs. }\label{fig:bicrystal}
\end{figure}

Consider a bicrystal  (see Fig.~\ref{fig:bicrystal}) that is periodic in the horizontal direction (along $x$) and bounded in the vertical direction (along $y$) by top and bottom surfaces.
We examine shear-coupled GB migration under free-surface and fixed-surface boundary conditions (BCs).
Different boundary conditions generate different stresses that may drive disconnection motion and hence affect  GB migration.
In other words, the effects of the BCs on the GB migration are embodied in  image stresses contained in the internal stress field $\bm{\sigma}$ in Eq.~\eqref{eq:v_d}.
To calculate the internal stress field, we first solve the appropriate elasticity problem in a bicrystal with prescribed BCs.

\begin{figure}[!b]
\centering
\includegraphics[width=1.\textwidth]{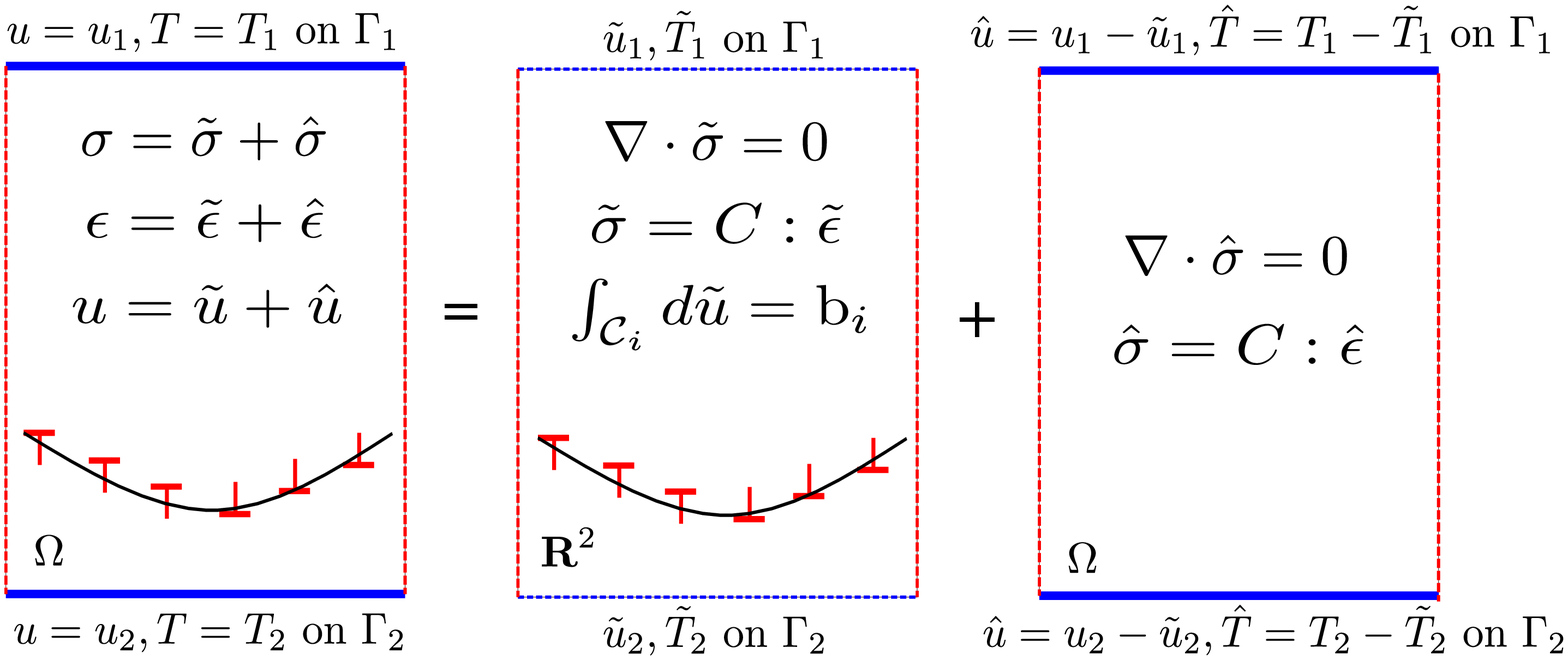}
\caption{Decomposition of the elasticity problem in a domain $\Omega$ containing disconnections with prescribed BC on $\Gamma_1$ and $\Gamma_2$ and periodic BCs on the left and right edges. The solution ($\sigma, \epsilon, u$) to the elasticity problem in $\Omega$ containing disconnections is the superposition of the fields ($\tilde{\sigma}, \tilde{\epsilon}, \tilde{u}$) generated by the disconnections in an infinite medium $\textbf{R}^2$, and the solution ($\hat{\sigma}, \hat{\epsilon}, \hat{u}$) to an elasticity problem in a disconnection-free domain $\Omega$ with complementary BCs on $\Gamma_1$ and $\Gamma_2$.}\label{fig:stress}
\end{figure}

We use the superposition approach previously employed by \cite{needleman1995discrete} to obtain the internal stress field.
Suppose the bicrystal domain $\Omega$ (with disconnections distributed along the GB) is a rectangle that extends periodically along the $x$-direction and is bounded by a top $\Gamma_1$ ($y=L_y$) and bottom surface $\Gamma_2$ (at $y=-L_y$); see Fig.~\ref{fig:stress}.
Consider the following BCs on the elastic displacements/surface tractions on $\Gamma_1$  and $\Gamma_2$:
\begin{equation} \label{eq:BCs}
\begin{aligned}
\bm{u} = \bm{u}_1 \quad \text{or} \quad \bm{T}=\bm{T}_1, \quad  \text{on $\Gamma_1$,} \\
\bm{u} = \bm{u}_2 \quad \text{or} \quad \bm{T}=\bm{T}_2, \quad  \text{on $\Gamma_2$.}
\end{aligned}
\end{equation}

Since we assume linear (isotropic) elasticity in the bicrystal domain ($\Omega$), application of the superposition principle allows us to decompose the  stress ($\bm{\sigma}$), strain ($\bm{\epsilon}$) and displacement ($\bm{u}$) fields into two sets as illustrated in Fig.~\ref{fig:stress}
\begin{eqnarray}
\bm{\sigma}=\tilde{\bm{\sigma}}+\hat{\bm{\sigma}}, \quad
\bm{\epsilon}=\tilde{\bm{\epsilon}}+\hat{\bm{\epsilon}}, \quad
\bm{u}=\tilde{\bm{u}}+\hat{\bm{u}} \quad \text{in $\Omega$.}
\end{eqnarray}
Here, the first set of fields ($\tilde{\bm{\sigma}},\tilde{\bm{\epsilon}},\tilde{\bm{u}}$) is generated by all of the disconnections in the infinite medium  and a second set of fields ($\hat{\bm{\sigma}},\hat{\bm{\epsilon}},\hat{\bm{u}}$)  corresponding to  image stresses (associated with the)  constraints at the top and bottom surfaces.


The fields ($\tilde{\bm{\sigma}},\tilde{\bm{\epsilon}},\tilde{\bm{u}}$) for a discrete dislocation in an infinite medium are well-known \citep[Chap.3]{hirth1982theory}.
However, since we consider periodic BCs in the horizontal direction, we must account for the fields of the replicas of this discrete dislocation in all periodic cells.
The analytical solutions for such a string of dislocations are also known \citep{hirth1982theory,needleman1995discrete}.

Let $\bm{\sigma_i}(x-x_0,y-y_0)$ be the stress at  point ($x,y$) generated by a disconnection of mode $i$ located at ($x_0,y_0$) (in the current cell) and its periodic replicas (see Eq.~(\ref{eq:period_soln2})).
Similarly,  the stress at point ($x,y$) from all disconnection modes along the GB  is now simply
\begin{eqnarray}
\tilde{\bm{\sigma}}(x,y)= \int_\mathcal{L}\sum_i\bm{\sigma_i}(x-x_0,y-y_0)\rho_i(x_0,y_0)ds,
\end{eqnarray}
where the integral over $s$ denotes locations  ($x_0,y_0$) along the GB, denoted by $\mathcal{L}$.
We can similarly obtain the corresponding strain and the displacement fields $(\tilde{\bm{\epsilon}},\tilde{\bm{u}})$.

The fields $(\tilde{\bm{\sigma}},\tilde{\bm{\epsilon}},\tilde{\bm{u}})$ are obtained for disconnections in an infinite medium by ignoring the boundary constraints at the top and bottom surfaces.
The associated displacements and surface tractions along the top and bottom surfaces $\Gamma_k$ ($\tilde{\bm{u}}_k,\tilde{\bm{T}}_k$), do not necessarily satisfy the prescribed BCs in Eq.~\eqref{eq:BCs}.

To ensure the total fields $(\bm{\sigma},\bm{\epsilon},\bm{u})$  satisfy the BCs, we introduce a second set of fields $(\hat{\bm{\sigma}},\hat{\bm{\epsilon}},\hat{\bm{u}})$ to be added to the original total fields that satisfy
\begin{eqnarray} \label{eq:equil_hat}
&&\nabla\cdot\hat{\bm{\sigma}}=\bm{0}, \quad
\hat{\bm{\sigma}}= \bm{C}:\hat{\bm{\epsilon}}, \quad
\hat{\bm{\epsilon}}=\frac{1}{2}[\nabla\hat{\bm{u}}+(\nabla\hat{\bm{u}})^{\textbf{T}}] \quad \text{in $\Omega$} \\
\label{eq:BC1_hat}
&& \hat{\bm{u}}=\bm{u}_1-\tilde{\bm{u}}_1 \quad \text{or} \quad \hat{\bm{T}}=\bm{T}_1-\tilde{\bm{T}}_1, \quad \text{on $\Gamma_1$} \\
\label{eq:BC2_hat}
&& \hat{\bm{u}}=\bm{u}_2-\tilde{\bm{u}}_2 \quad \text{or} \quad \hat{\bm{T}}=\bm{T}_2-\tilde{\bm{T}}_2, \quad \text{on $\Gamma_2$}.
\end{eqnarray}
This is an elasticity problem in a disconnection-free domain subject to the complementary boundary conditions (the difference between the values of the prescribed BCs and the boundary values from the original set of fields).
This elasticity problem has analytical solutions for a rectangle domain under both  free- and fixed-surface BCs.

For  free-surface BCs, the tractions along the surfaces are zero ($\bm{T}_1=\bm{T}_2=\bm{0}$. 
We  solve for the stress field $\hat{\bm{\sigma}}$ using an Airy stress function and Fourier analysis .
The solutions for the Airy function and stress are given in  Eqs.~(\ref{eq:soln_free2}--\ref{eq:airy_fun}).

For the fixed-surface BCs, we set the total displacements to be zero.
As shown in Fig.~\ref{fig:bicrystal}, in addition to the elastic displacement $\bm{u}$, the total displacement $\bm{D}$ includes the ``plastic'' displacement $\bm{B}$ associated with grain translation, i.e., $\bm{D}=\bm{u}+\bm{B}$.
Note that the grain translation Eq.~\eqref{eq:translation} is the relative lateral ``plastic'' displacement between the top and bottom surfaces.
If we define, without loss of generality, a positive disconnection as having a Burgers vector in the $+x$ direction for all disconnection modes ($b_i>0$ for all $i$), we may write the lateral ``plastic'' displacements for the top and bottom surfaces separately as
\begin{equation}\label{eq:B}
\begin{aligned}
B_1 = \int \sum_i v_i (\rho_i^+ + c_i)b_i\ dt \quad \text{on $\Gamma_1$,}\\
B_2 = \int \sum_i v_i (\rho_i^- - c_i)b_i\ dt\quad \text{on $\Gamma_2$,}
\end{aligned}
\end{equation}
where
\begin{eqnarray}
\rho_i^+ = \max(\rho_i,0), \quad \rho_i^- = \min(\rho_i,0).
\end{eqnarray}
To satisfy the fixed-surface BCs in Eqs.~(\ref{eq:BC1_hat}--\ref{eq:BC2_hat}), we  set $\bm{u}_1=-\bm{B}_1=(-B_1,0)$ and $\bm{u}_2=-\bm{B}_2=(-B_2,0)$.
We solve this elastic boundary value problem for the displacement field $\hat{\bm{u}}$ by employing a displacement formulation of the equilibrium equation (Eq.~\eqref{eq:equil_hat}) such that $\hat{\bm{u}}$ satisfies the biharmonic equation when the body force is zero; see Eq.~(\ref{eq:soln_fix2}--\ref{eq:soln_fix3}).
The stress field $\hat{\bm{\sigma}}$ is then obtained from the linear elastic constitutive equations.

By applying the superposition approach described above, we determine the internal stress in the  GB equation of motion Eq.~(\ref{eq:multiEOM}) that includes the effects of the boundary constraints.
We now examine the effects of such  constraints on  GB migration by applying our continuum model to GB migration  in a bicrystal under  fixed- and free-surface BCs.

\section{SIMULATION RESULTS}
\label{sec:simulations}
We apply our continuum model in a series of numerical simulations to study shear-coupled GB migration with multiple disconnection modes as a function of  boundary constraints,  types of driving forces and temperature.
To parameterize this model, we employ the atomistic simulation data for copper using an embedded atom interatomic potential  \citep{Mishin2001Cu}: shear modulus $\mu=45$ GPa, poisson ratio $\nu=0.36$, and lattice constant $a_0=3.615$~\AA.
\cite{cahn2006coupling} determine the grain boundary energy densities of several [001] symmetric tilt grain boundaries with this potential: for $\Sigma$13 (015) $\gamma=0.878$~J/m$^2$ and for $\Sigma$37 (057) $\gamma=0.732$~J/m$^2$.
We also use the following parameters: the disconnection mobility $M_d=100$~m$^2$/(J$\cdot$s) and the length of the GB in the bicrystal $L_0=100$~\AA.

\subsection{Grain Boundary Migration in a Bicrystal}
\label{subsec:shear_couple}
We first study GB migration in a bicrystal driven by a jump in chemical potential across the GB (i.e., a synthetic driving force \citep{janssens2006computing}) under free- and fixed-surface BCs at the top and bottom surfaces.
Molecular dynamics simulations were previously employed by \cite{thomas2017reconciling} to investigate  shear-coupled migration of a flat GB.
When free-surface BCs were employed, they observed that GBs readily migrate through the entire bicrystal and the  grains show lateral translation with respect to one another, as expected during low temperature, shear-coupled GB migration.
However, when fixed-surface BCs were employed at low temperature, the GB  migrates a short distance then stagnates.
They attributed this stagnation to stress generation during shear-coupled migration with fixed-surface BCs.
On the other hand, when fixed-surface BCs were employed at a higher temperature, the GB migrates through the entire bicrystal with no net lateral translation of the two grains.
They attributed this change in behavior with temperature to the thermal activation of secondary mode disconnections  ($\beta>0$) with opposite sign shear coupling as compared with that resulting from the primary disconnection mode ($\beta<0$).

\begin{figure}[!t]
\centering
\subfloat[]{\includegraphics[width=0.33\textwidth]{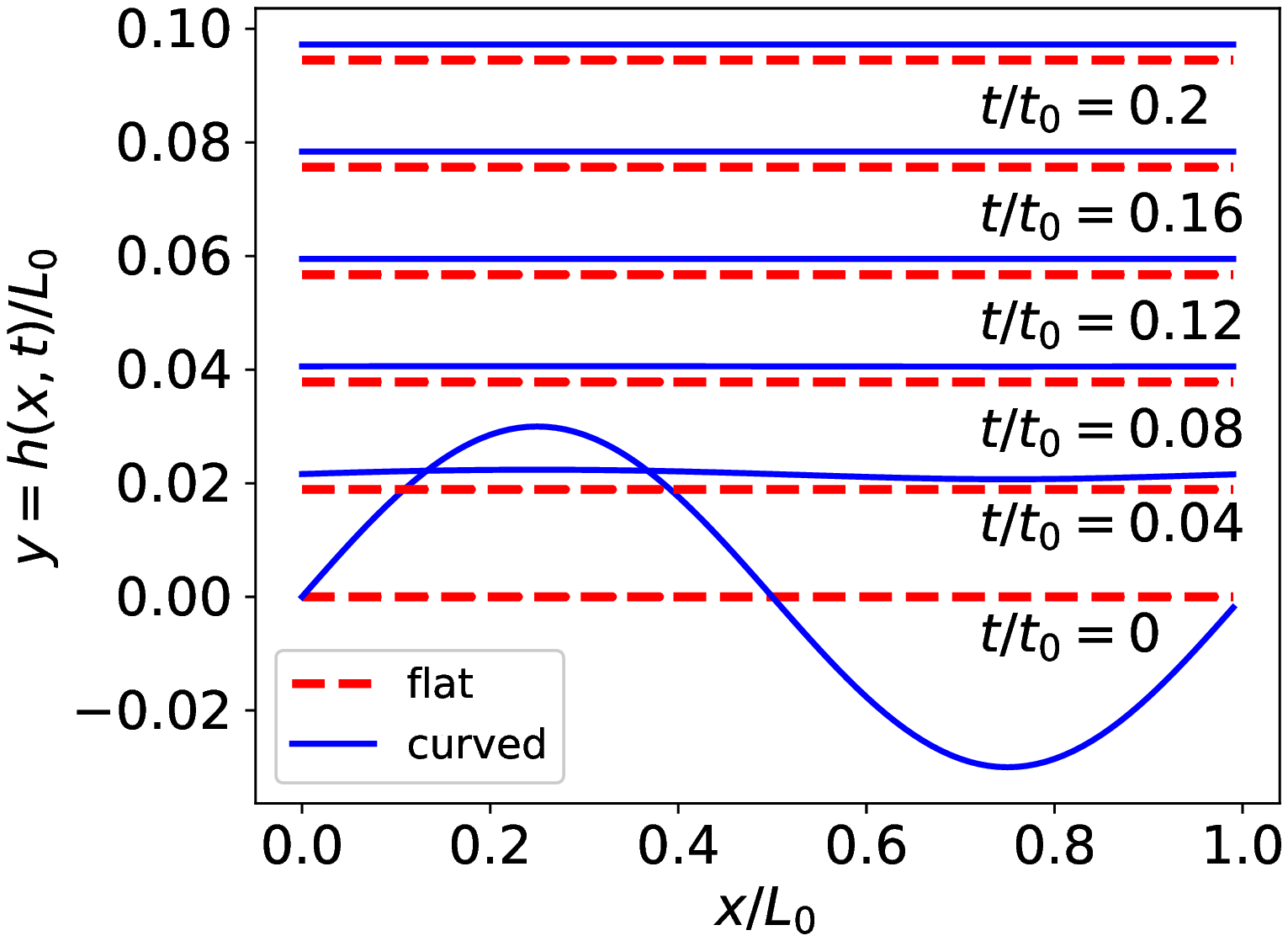}}
\subfloat[]{\includegraphics[width=0.33\textwidth]{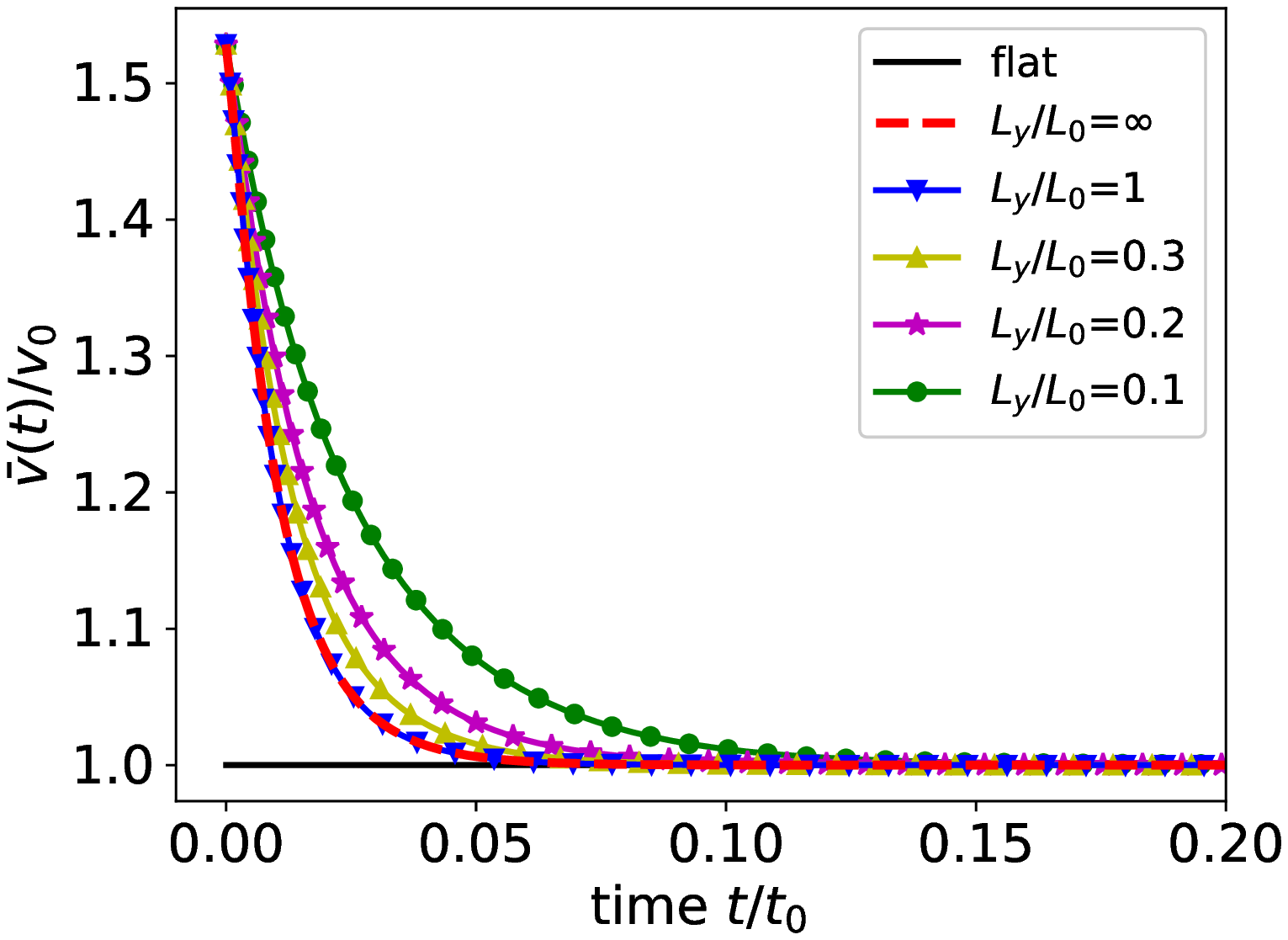}}
\subfloat[]{\includegraphics[width=0.33\textwidth]{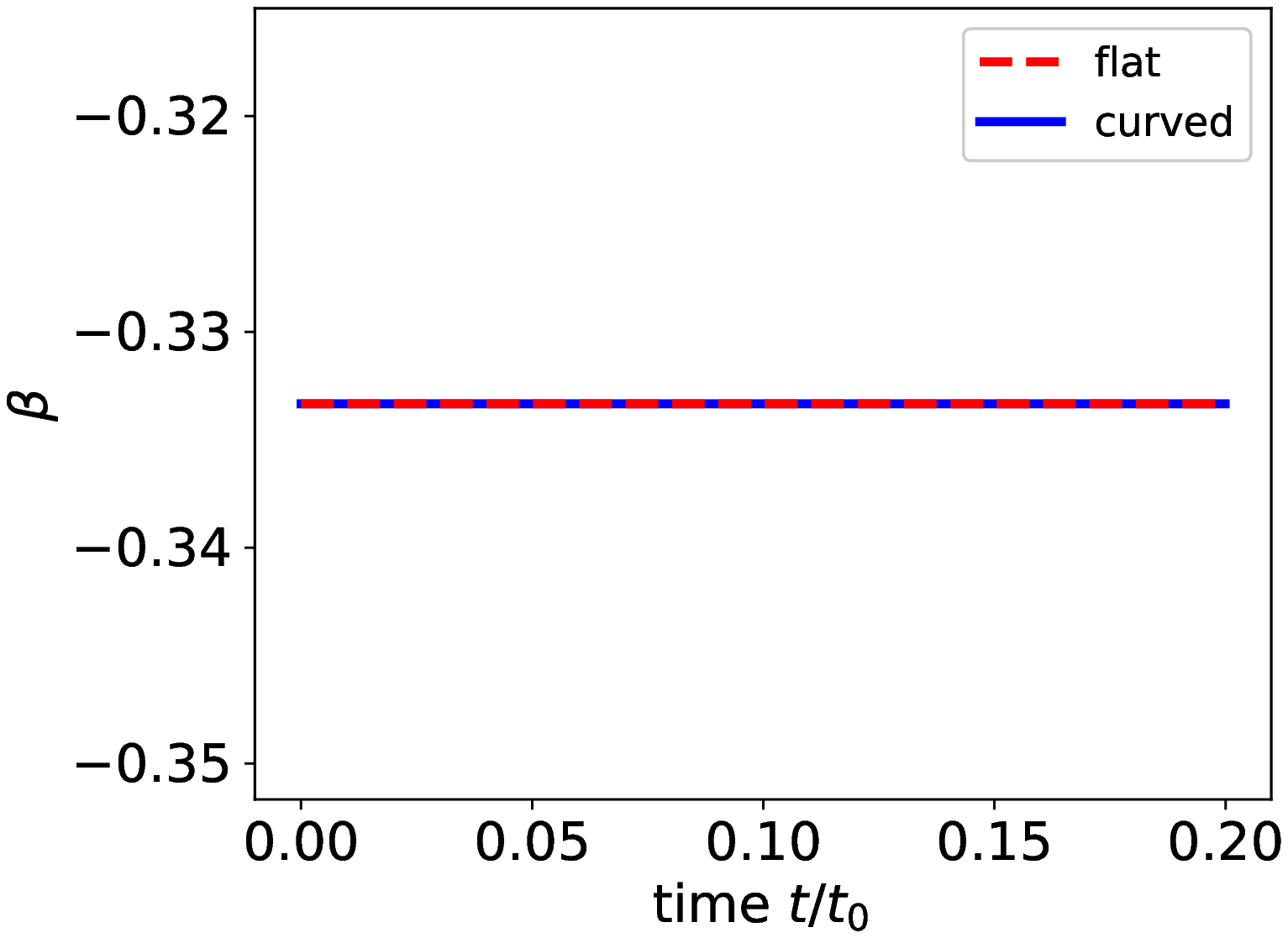}}
\caption{Single-mode migration of a $\Sigma$37 GB with free-surface BCs at the top and bottom surfaces ($y$=$\pm L_0$).
(a) The migration of an initially flat (red dashed curves) and sinusoidal (solid blue curves) GB under a chemical potential jump driving force $\Psi$=$-1$ meV/\AA$^3$. The GBs migrate until they reach the top surface.
(b) The average GB velocities $\bar{v}(t)/v_0$ of the initially flat and sinusoidal GBs for several bicrystal lengths in the $y$-direction , i.e., $L_y/L_0=0.1, 0.2, 0.3, 1, \infty$.
$v_0$ is the velocity of the flat GB.
(c) The coupling factor of the flat and curved GBs during migration.
}\label{fig:single_free}
\end{figure}

The continuum model enables us to easily switch on/off different disconnection modes to ascertain how GB migration is affected by changing BC-type at low temperature and  changing  temperature for the fixed-surface BC case.
We consider $\Sigma$37 symmetric tilt GB migration based on a single disconnection mode $(a_0/\sqrt{74},3a_0/\sqrt{74})$ for two types of BCs.
The GB migration results are shown in Fig.~\ref{fig:single_free}(a) for the free-surface BC case for both an initially flat and sinusoidal GB.
We  see that both GBs readily migrate to the end of the bicrystal, driven by a chemical potential jump  ($\Psi = -1$ meV/\AA$^3$).
To show the effect of initial GB profile, we examine the average GB migration velocity $\bar{v}(t)=\langle v(x,t)\rangle_x$ (i.e., averaged over position, $x$ in Fig.~\ref{fig:single_free}(a)) as a function of time; see Fig.~\ref{fig:single_free}(b).
While the flat GB migrates at a constant velocity $v_0$, the initially sinusoidal GB initially migrates at a larger mean velocity but slows with increasing time as the GB flattens during migration (the initial rate of migration is associated with curved GBs).
The difference between the initial migration rates of the flat and sinusoidal GB profiles is attributed to the initial active disconnections along the GB.
Both positive and negative disconnections actively glide along the GB (in opposite directions) and lead to GB migration (in the same direction) that results in a non-vanishing average migration velocity, as suggested by the term $|\rho|$ in the GB  equation of motion, Eq.~\ref{eq:multiEOM}.
While the flat GB has zero local disconnection density everywhere ($\rho=0$), the sinusoidal GB has a spatially varying disconnection density ($\rho\neq 0$); this implies the sinusoidal GB has more active disconnections and hence a larger initial migration rate.
The initially sinusoidal GB slows as it migrates and flattens as the original excess disconnection density relaxes to its steady-state value under the combined driving force of both the elastic interactions between disconnections and capillarity. As the GB approaches the free surface, the stress field in the bicrystal associated with the zero traction BC slows the flattening.
This effect may be understood in terms of the zero traction BC producing effective image disconnections on the other side of the free surface.
This image stress field cancels the traction at the free surface and exerts a back stress on the disconnections (associated with the disconnection Burgers vector) along the GB through its associated Peach-Koehler force that prevents disconnection glide associated with GB flattening.
This free surface (zero traction or image disconnection) effect is verified by changing the length of the bicrystal (i.e., the simulation cell width in the direction normal to the GB, i.e., $L_y$), as shown in Fig.~\ref{fig:single_free}(b).
Decreasing $L_y$ implies slower relaxation of the mean GB velocity.
In Fig.~\ref{fig:single_free}(c), we verify that the coupling factor of the flat and curved GBs (for every point along the GB) is equal to the coupling factor of the primary disconnection mode, $\beta = -H_{10}/b_{10}=-1/3$.

\begin{figure}[!t]
\centering
\subfloat[]{\includegraphics[width=0.4\textwidth]{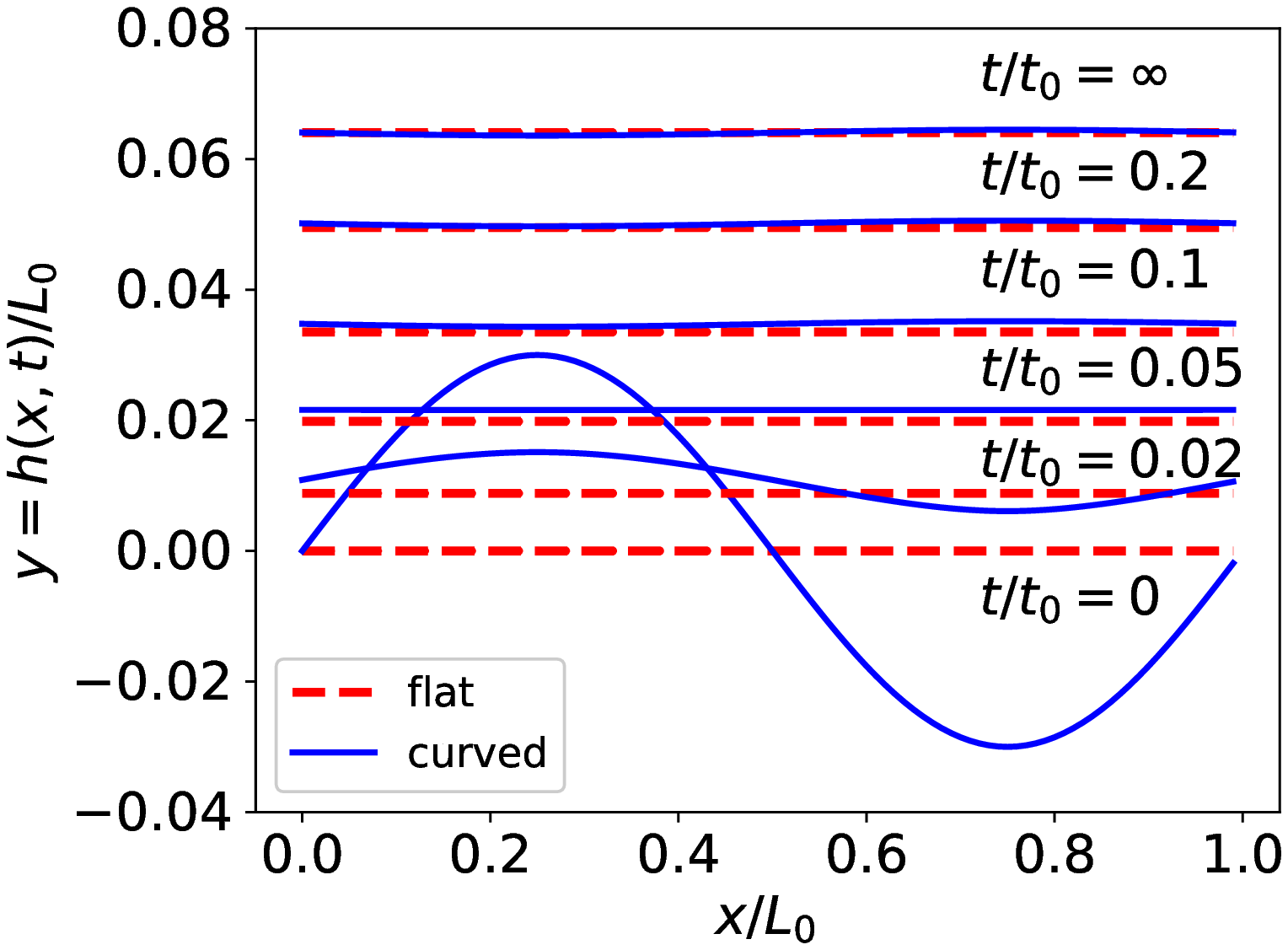}}
\subfloat[]{\includegraphics[width=0.4\textwidth]{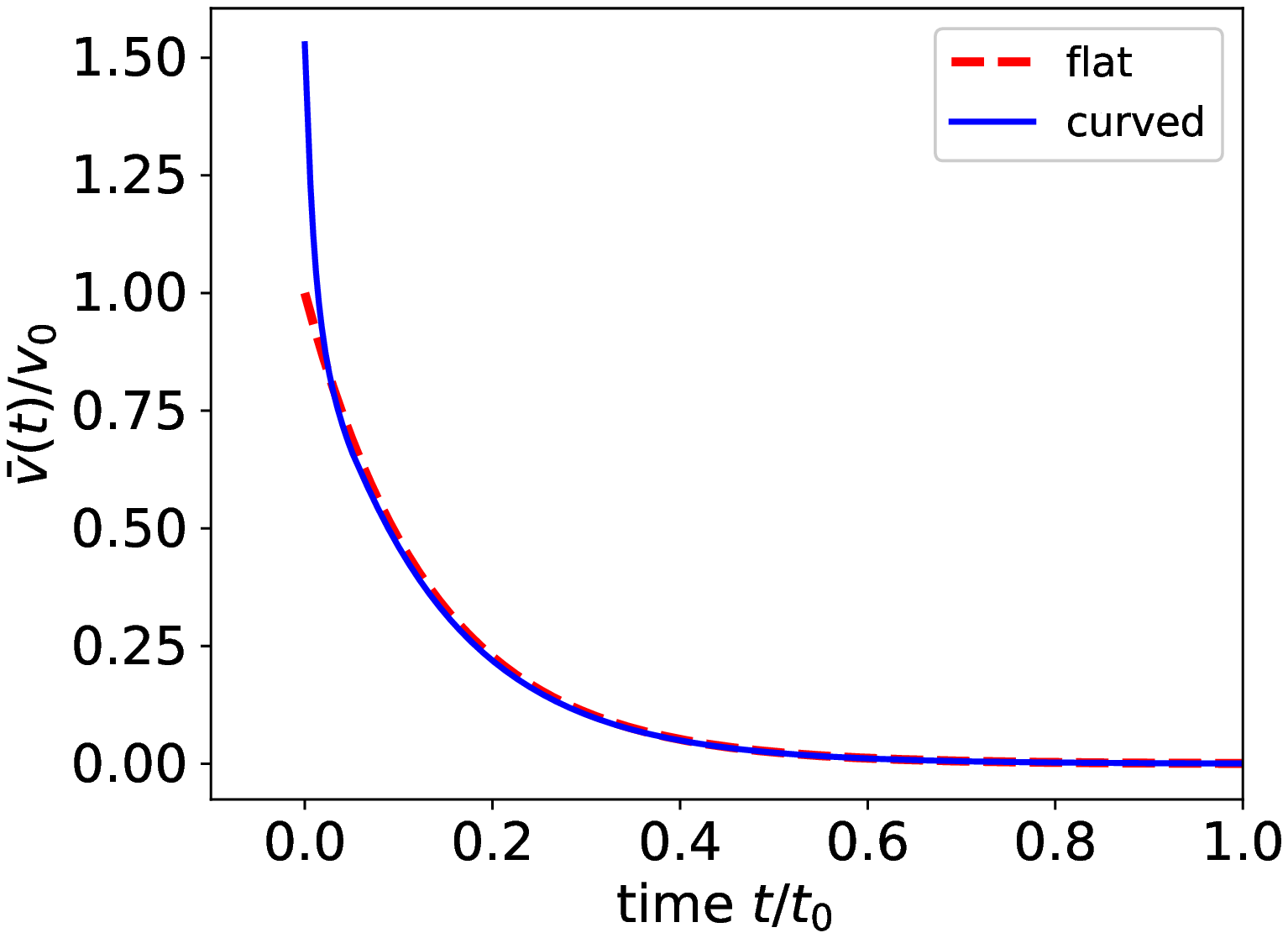}}\\
\subfloat[]{\includegraphics[width=0.4\textwidth]{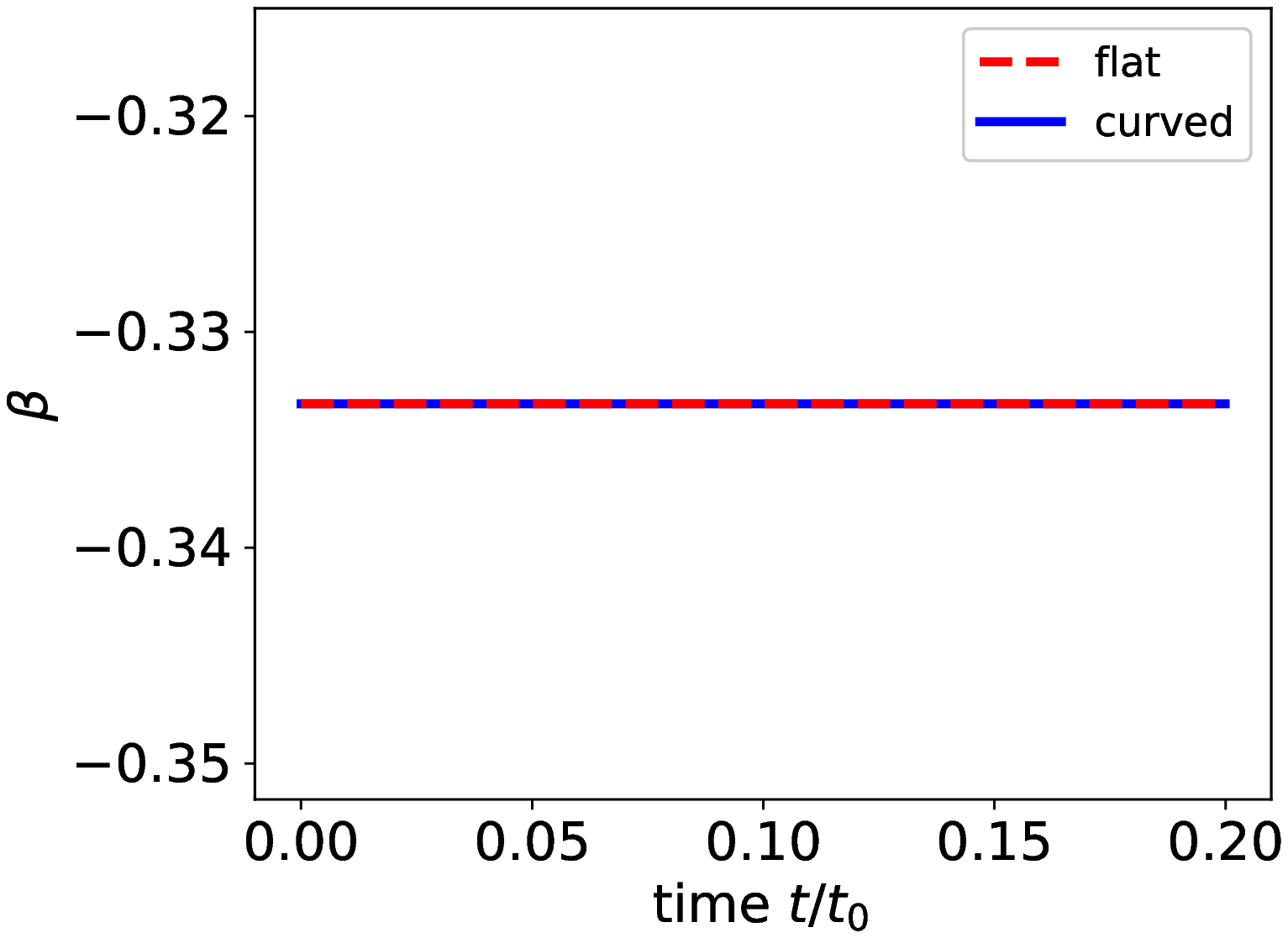}}
\subfloat[]{\includegraphics[width=0.4\textwidth]{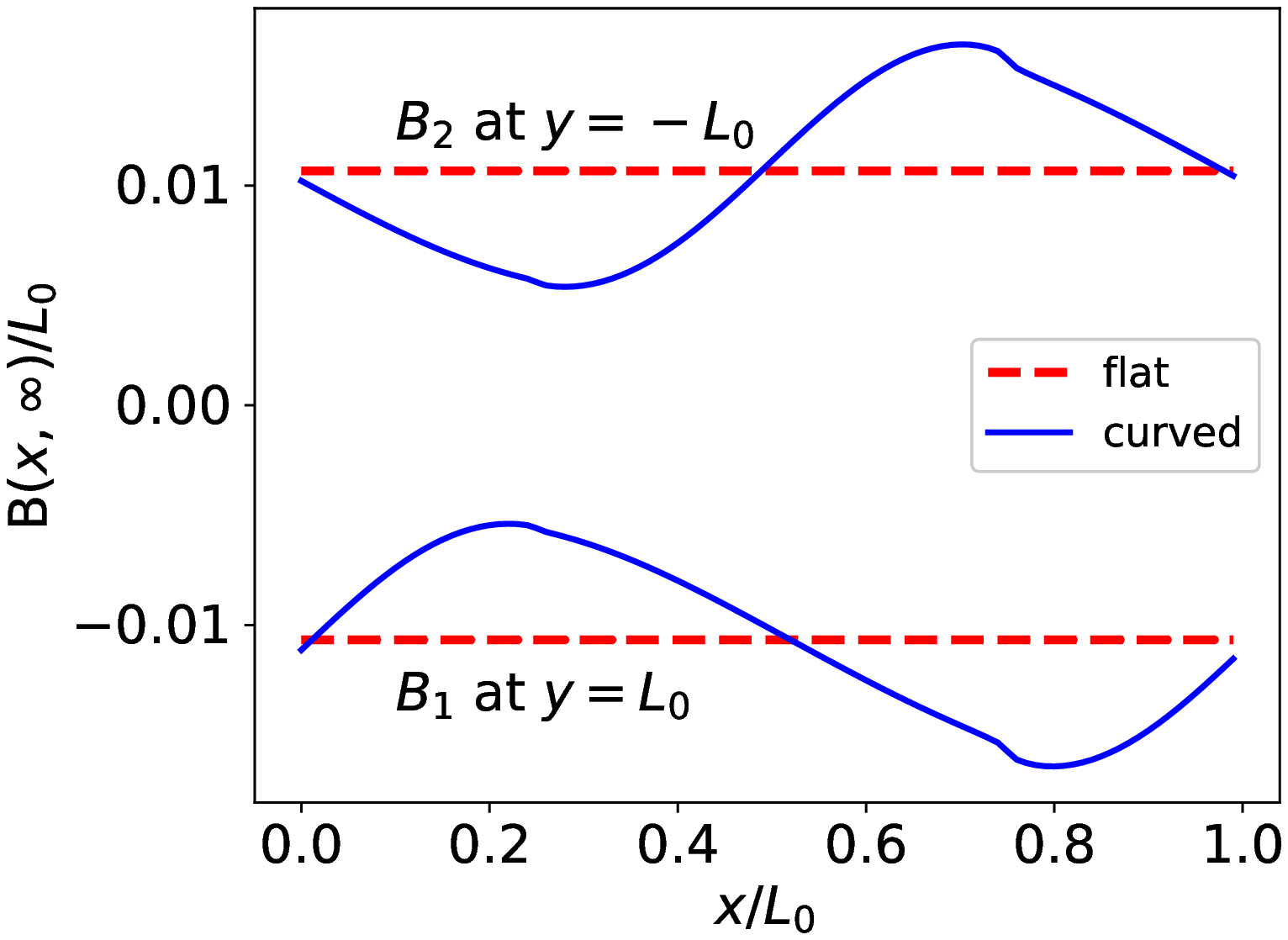}}
\caption{Single-mode migration of a $\Sigma$37 GB with fixed-surface BCs at the top and bottom boundaries ($y$=$\pm L_0$).
(a) The migration of a flat  and curved GB under a chemical potential jump driving force $\Psi$=$-1$ meV/\AA$^3$.
The GBs migrate a short distance and then stagnate.
(b) The average GB velocities $\bar{v}(t)$ of the flat and curved GBs.
(c) The coupling factor of the flat and curved GBs during migration.
(d) The coupled shear deformation at the top and bottom surfaces $B(x,t=\infty)$ (see Eq.~\ref{eq:B}) when GB migration stagnates.
}
\label{fig:single_fixed}
\end{figure}

\begin{figure}[!ht]
\centering
\subfloat[]{\includegraphics[width=0.4\textwidth]{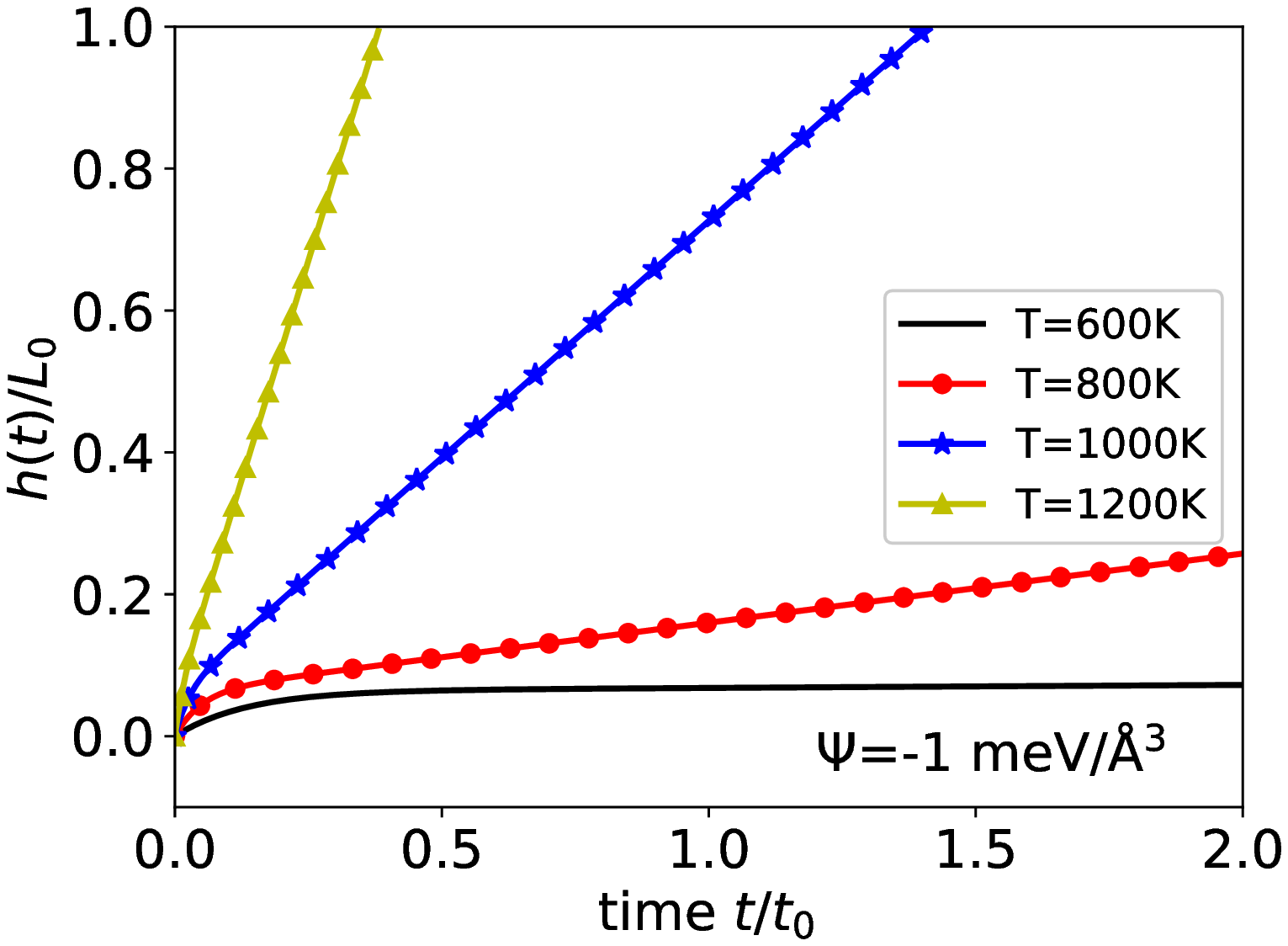}}
\subfloat[]{\includegraphics[width=0.4\textwidth]{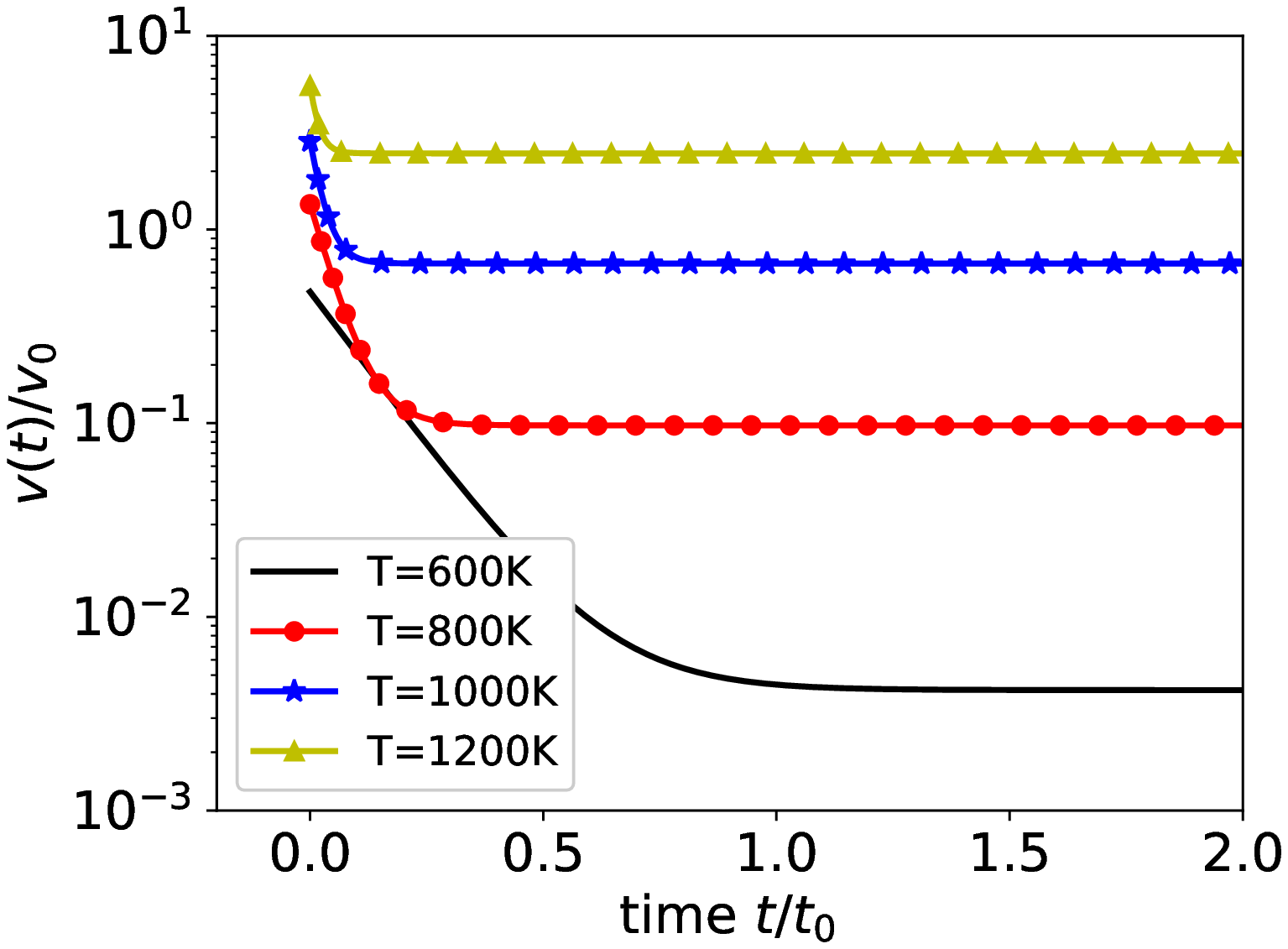}}\\
\subfloat[]{\includegraphics[width=0.4\textwidth]{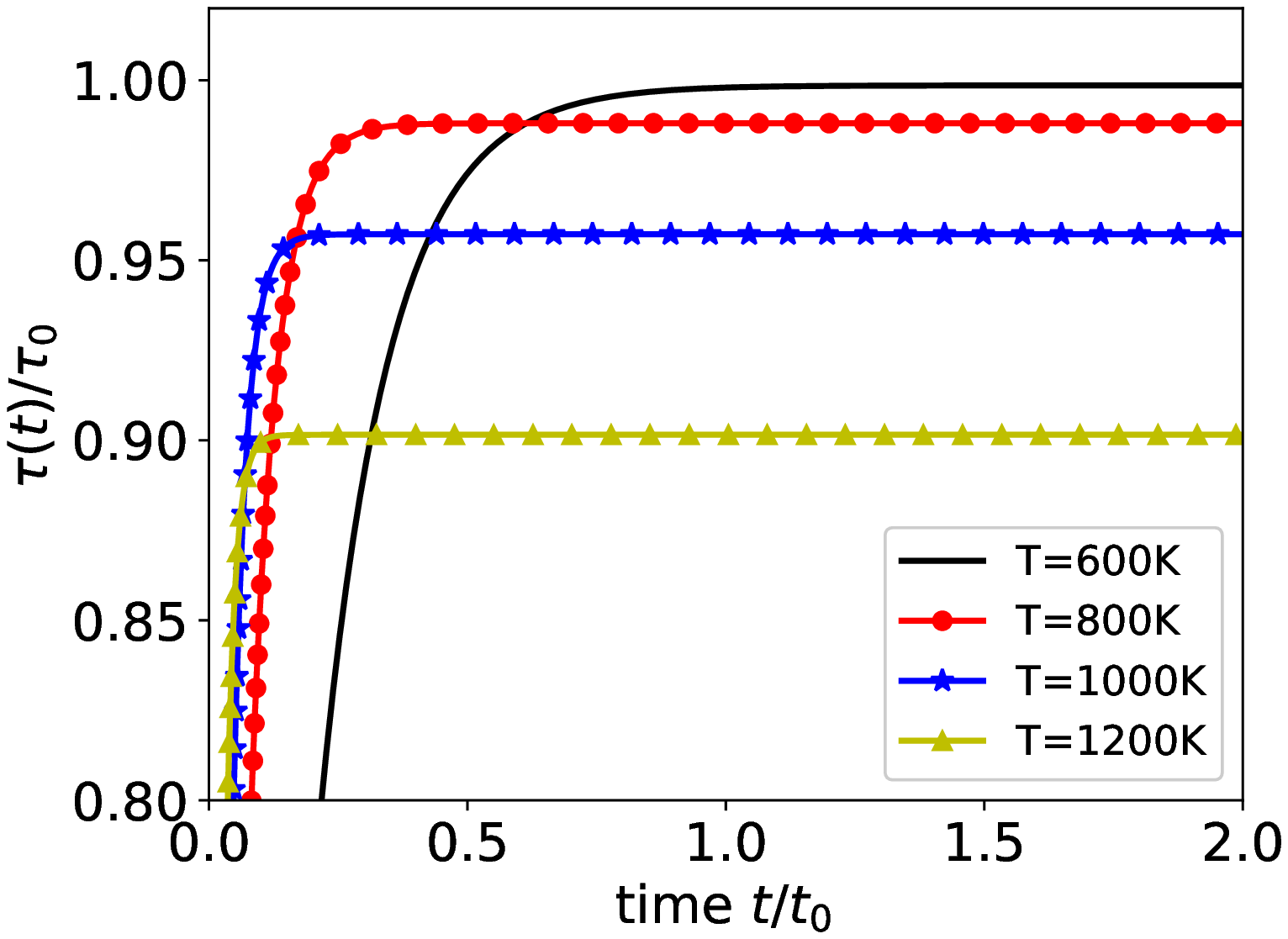}}
\subfloat[]{\includegraphics[width=0.4\textwidth]{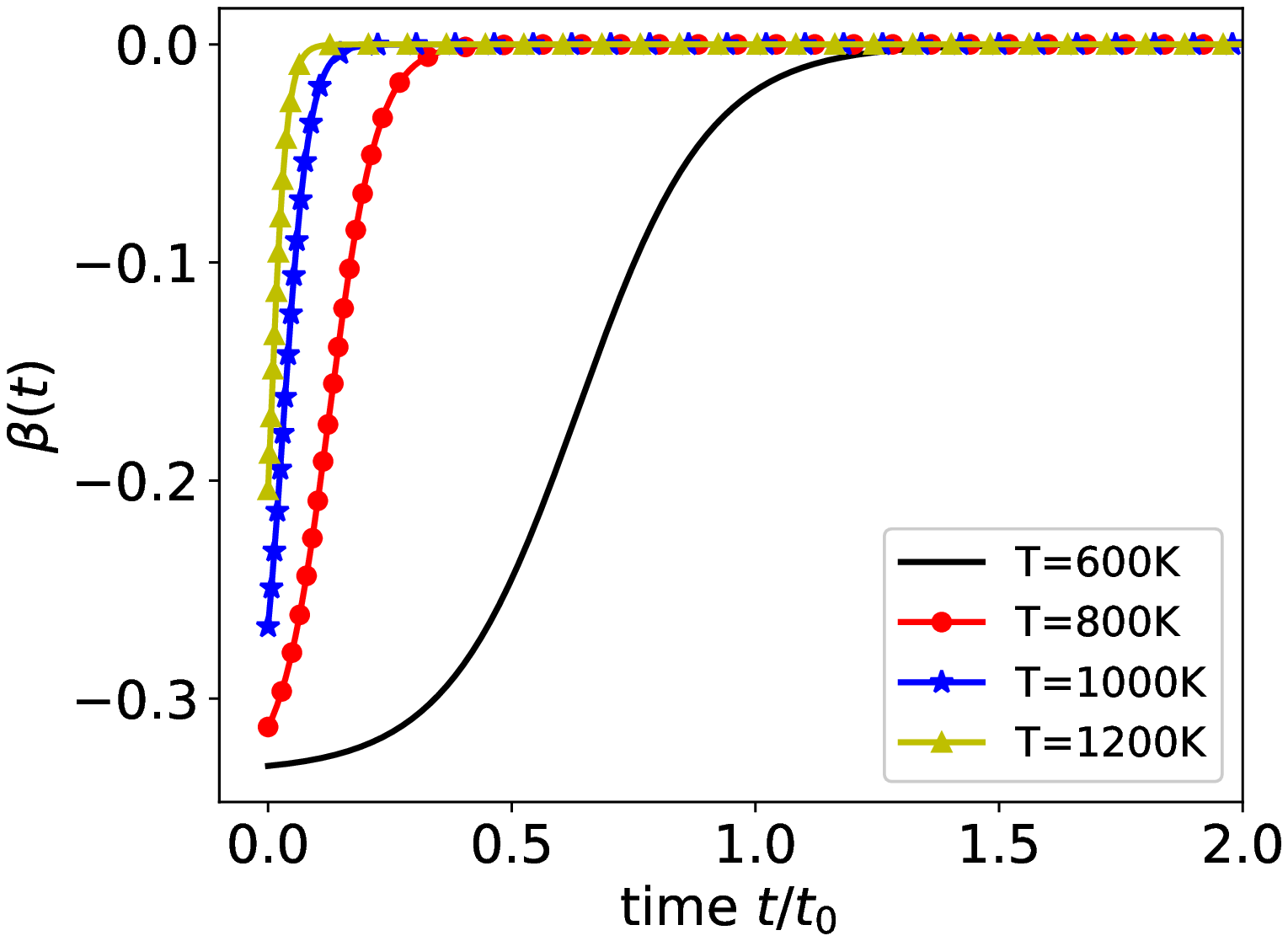}}
\caption{Multi-mode migration of the $\Sigma$37 GB under fixed-surface boundary condition as a function of temperatures.
(a) The GB position $h(t)/L_0$ as a function of time for $T=600$, 800, 1000, 1200~K.
The  GB reaches the top boundary at $h/L_0=1$.
(b) The GB migration velocity $v(t)/v_0$, where $v_0=L_0/t_0$.  Following an initial transient, the GB velocity approaches a temperature-dependent steady-state velocity.
(c) The time dependence of the image shear stress $\tau(t)/\tau_0$ in the bicrystal (associated with the fixed-surface boundary conditions).  Following an initial transient, the image stress approaches a temperature-dependent  value such consistent with  zero lateral grain translation.
(d) The time dependence of the shear coupling factor $\beta(t)$.
$\beta(0)$ is temperature-dependent and $\beta(t)\rightarrow0 $ at late time, where the rate of lateral grain translation tends to zero in steady-state GB migration.
}\label{fig:multi_fixed}
\end{figure}

Figure~\ref{fig:single_fixed}(a) shows the migration of both initially flat and sinusoidal GBs under a chemical potential jump driving force  ($\Psi = -1$ meV/\AA$^3$) for the fixed-surface BCs for bicrystal length $L_y=L_0$.
Like in the MD simulations \citep{thomas2017reconciling}, the GBs migrate a short distance and then stagnate at  $y/L_0\approx6\times 10^{-2}$.
Unlike in the free-surface BC case (where GBs can migrate to the end of the bicrystal cell), the lateral translation of one grain relative to the other during migration is constrained here by the fixed-surface BCs.
Hence, as the GB migrates, the stress in the bicrystal accumulates which, in turn, increases the Peach-Koehler force on the disconnections that opposes the driving force associated with the chemical potential jump and hence the GB motion slows, as seen in  Fig.~\ref{fig:single_fixed}(b).
When the two forces balance, the GB stagnates (see Fig.~\ref{fig:multi_fixed}(c)).
Moreover, when the GB migration stagnates, the curved GB induces nonuniform shear along the GB (see Fig.~\ref{fig:single_fixed}(d)) and therefore remains (slightly) curved.
(The small jogs in $B(x,\infty)$ in this plot are associated with the absolute value of $\rho$ in the disconnection flux, i.e., Eq.~(\ref{eq:flux}).)
Again, we verify  that the coupling factor for GB migration for both the flat and curved GBs is  $\beta=-1/3$ (see  Fig.~\ref{fig:single_fixed}(c)).

We now examine  shear-coupled GB migration for a flat GB where we allow for the possibility of  multiple disconnection modes (multi-mode GB migration) with fixed-surface BCs (at $y=\pm L_0$).
We drive the migration of a $\Sigma37$ GB with a chemical potential jump  driving force at different temperatures and allow for all of the  disconnection modes shown in Figure~\ref{fig:dis_energy}(c) where the activation energies for disconnection nucleations are as shown in Fig.~\ref{fig:dis_energy}(a).
Figure~\ref{fig:multi_fixed}(a) shows that GB migration effectively stagnates at low temperature $T=600$~K (note this GB does migrate, albeit extremely slowly), but at higher temperatures it migrates at a constant velocity (following an initial transient) that increases with increasing temperature (see Fig.~\ref{fig:multi_fixed}(b)).
These observations may be interpreted as follows.
When the temperature is  small, only one disconnection mode is activated and GB motion will stagnate, as shown in Fig.~\ref{fig:single_fixed}.
With increasing temperature, the generation of disconnections of higher modes are thermally activated.
Some of these higher modes have $\beta$ with the opposite sign of the primary mode, allowing GB migration in the same direction as from the primary mode but with lateral translation in the opposite direction as the primary mode.
This means that the motion of the disconnections of the primary mode shears the bicrystal in one direction and the secondary (and perhaps other) mode disconnections unshear the bicrystal.
The Peach-Koehler force associated with the stress accumulated during GB migration ($\hat{\sigma}$) (arising from the fixed-surface BC during  shear-coupled GB migration) opposes the motion of primary mode disconnections but enhances the motion of secondary (and other) mode disconnections.
Migration under the action of these multiple modes with fixed-surface BCs is balanced such that continuing GB migration creates no additional shear deformation.

In the present case (flat GBs with fixed-surface BCs), steady-state GB motion is achieved when the lateral translation rates of the two grains is zero, $\partial_t B=0$.
In steady-state, there may be a non-zero  image stress $\hat{\sigma}$ in the bicrystal, that acts like an applied, external stress on the GB, as seen in Fig.~\ref{fig:multi_fixed}(c).
This steady-state stress is $\tau(t=\infty)=-\sum_i \Psi H_ib_ic_i/\sum_i b_i^2c_i$ (for a flat GB driven by a chemical potential jump).
When only one disconnection mode  operates (as in Fig.~\ref{fig:single_fixed}), this steady-state stress $\tau_0=-\Psi H_{10}/b_{1}$ gives rise to a Peach-Koehler force that cancels the chemical potential jump driving force and  GB migration stagnates.
When multiple disconnection modes are operating,  GB migration achieves a constant velocity which  decouples from the lateral grain translation; this implies that the shear-coupling factor $\beta(t)\rightarrow 0$ in Fig.~\ref{fig:multi_fixed}(d), i.e., pure migration and no net shear coupling.
While the MD simulations  \citep{thomas2017reconciling} showed   that the bicrystal alternately shears and unshears during GB migration, in the  continuum limit no such alternating shearing and unshearing occurs since disconnection nucleation is continuous.
The same approach can be applied to understand the initial ($t\rightarrow0$) shear-coupling factor.
The initial shear-coupling factor accounts for the nucleation and motion of disconnection of all modes, i.e., $\beta(t\rightarrow0)=\sum_i (\tau_0 b_i+\Psi H_i)c_ib_i/\sum_i(\tau_0 b_i+\Psi H_i)c_iH_i$.

With free-surface BCs, a flat GB with multiple disconnection modes will  migrate at a constant rate as it does in the single disconnection mode case, albeit with a different (temperature-dependent) velocity and with a different  (temperature-dependent) shear-coupling factor that represent appropriate averages of all of the disconnection modes.
The coupling factor is given by the same expression as for the initial coupling factor of GB migration under fixed-surface BCs.

The migration of curved (sinusoidal) GBs with multiple disconnection modes is influenced by both the effects of the BCs and the competition between different disconnection modes, as in the flat GB case.
Moreover, the shear-coupled migration behavior of curved GBs also strongly depends on the initial distribution of  density of disconnection of each mode.
Unlike in the single-mode case, however, the relationship between GB shape and the densities of disconnections of different modes is not unique in the multi-mode case.
In fact, for a  GB of a given shape $h(x)$, there are infinitely many possible disconnection density distributions that satisfy the GB shape constraint  $\partial_x h = \sum_i \rho_iH_i$.
In general, the disconnection density distribution is history-dependent.
In the next section, we  discuss  the equilibrium disconnection density profile for GBs with pinned ends.

\subsection{Evolution and Equilibrium of Grain Boundaries with Pinned Ends}
\label{subsec:GB_equil}
Unlike in the ideal case of bicrystals, in polycrystals, GBs are of finite length (area) - inevitably delimited by the triple junctions and higher-order junctions at (along) which multiple grains/GBs meet.
Not surprisingly, GB motion is affected by the resulting finite-size constraint and/or by the TJ dynamics.
We now consider the effects of finite GB lengths (areas) on the evolution of GBs; in particular, we  assume (for now) that the TJs are pinned.
To this end, we  first establish appropriate BCs at the two ends of a GB in Eqs.~(\ref{eq:migration}--\ref{eq:evolution}).
Fixed ends imply that $\partial_t h(0,t)=\partial_t h(L_0,t)=0$, where now $L_0$ is the distance between TJs.
We also assume that no disconnection flows through the TJs, i.e., $J_i(0,t)=J_i(L_0,t)=0$ for all disconnection modes $i$.
We now examine the  dynamics of such delimited GBs and their equilibrium  profiles (i.e., GB shape and disconnection density distribution) for different types of driving forces.
In all simulations, we assume that the initial GB is flat, $h(x,0)=0$, and there is zero net Burgers vector density $\rho_i(x,0)=0$ for all $i$.

\begin{figure}[!ht]
\centering
\subfloat[]{\includegraphics[width=0.4\textwidth]{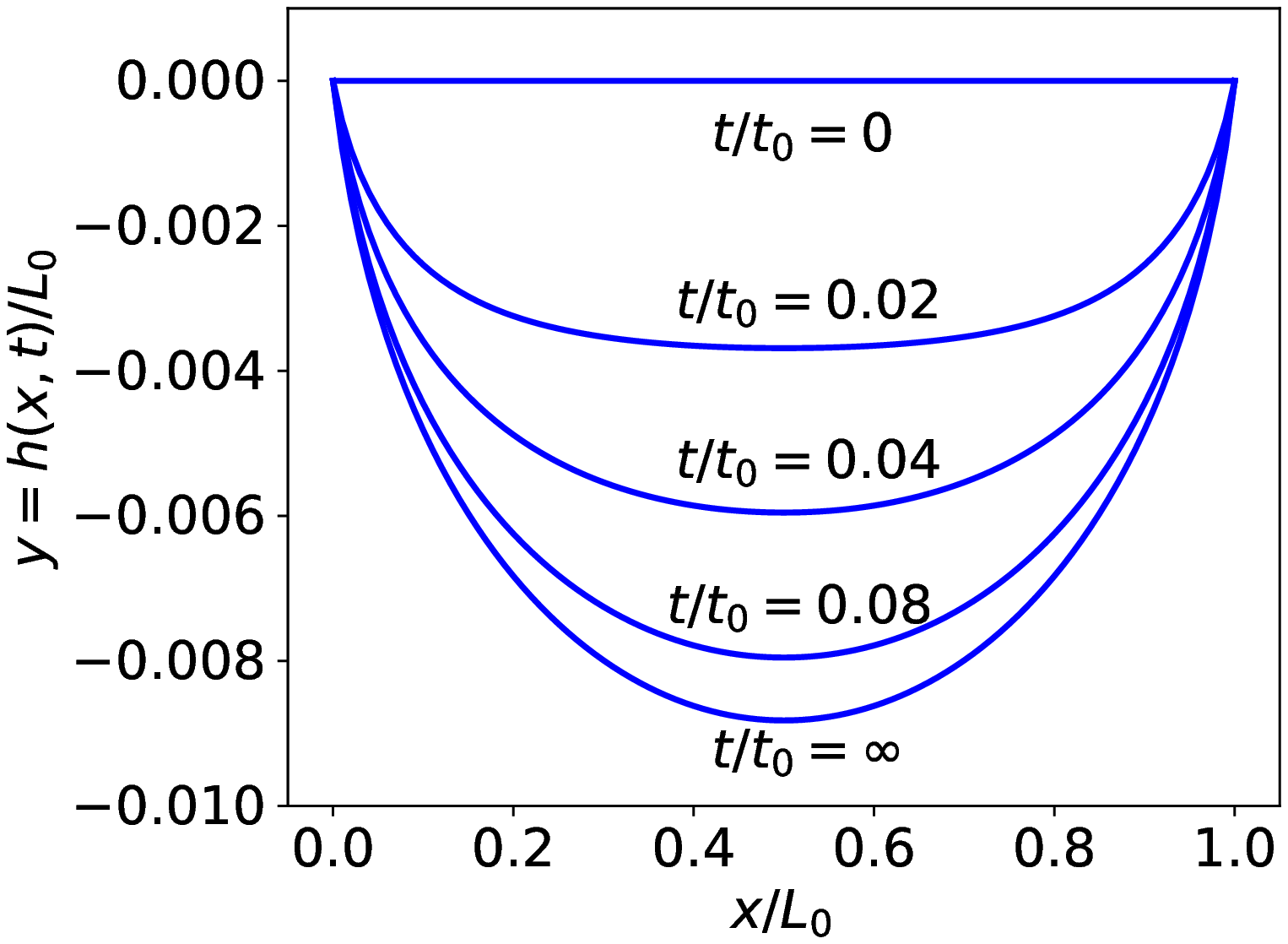}}
\subfloat[]{\includegraphics[width=0.4\textwidth]{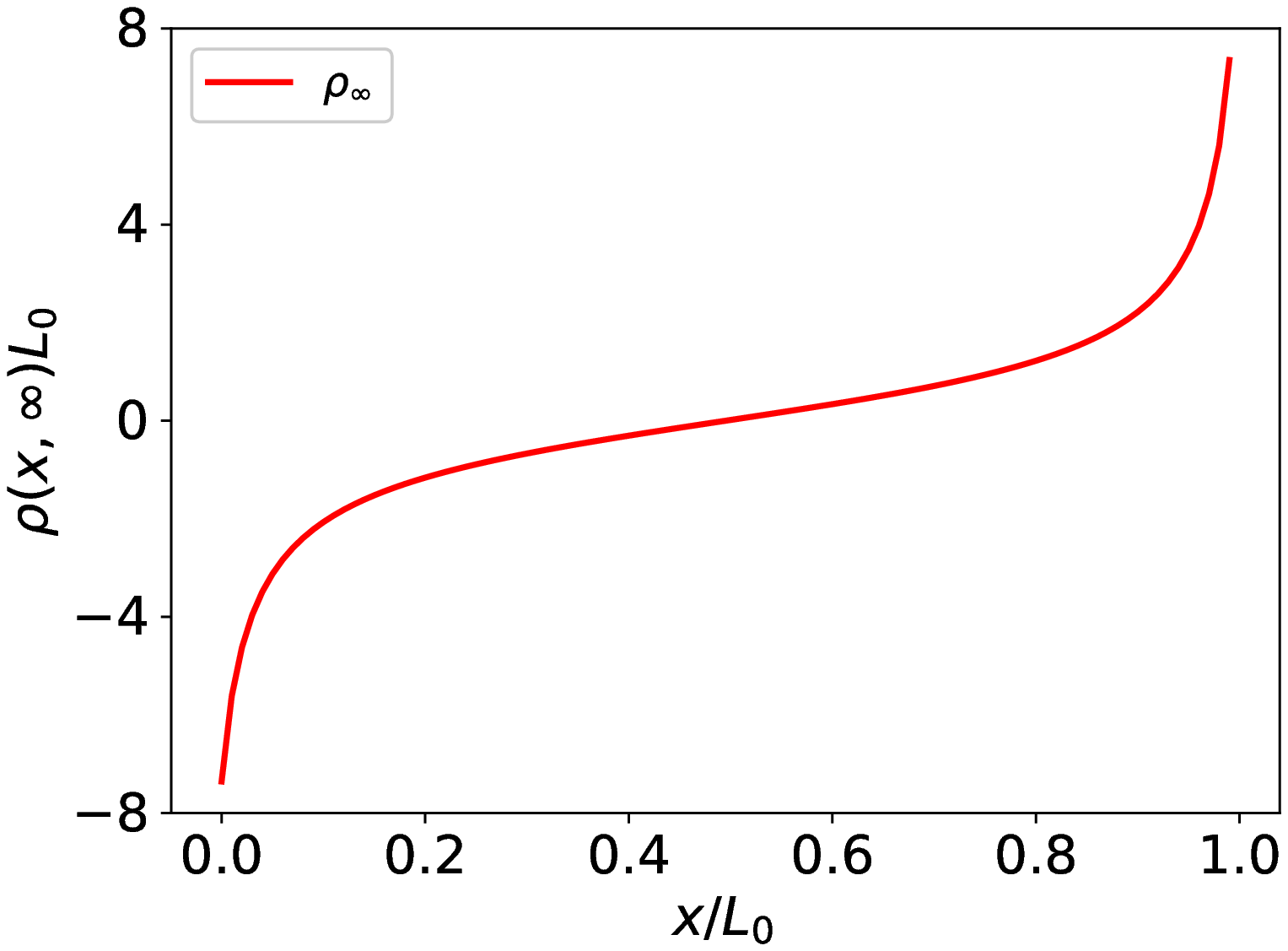}}
\caption{Single-mode GB migration of a $\Sigma37$ GB with pinned ends under an applied stress.
(a) The evolution of the $\Sigma$37 GB driven by applied stress $\tau_0 = 5\times10^{-3}\mu$ at $T=600$~K.
(b) The disconnection density distribution in equilibrium, $\rho(x,t=\infty)$.}
\label{fig:single_tau}
\end{figure}

We first consider the GB migration of a $\Sigma37$ GB with only a single disconnection mode under an applied shear stress $\tau_0/\mu = 5\times10^{-3}$, as shown in Fig.~\ref{fig:single_tau}.
The initially flat GB bows out and the disconnections pile up near the two, pinned ends of the GB.
At late time, the GB approaches a steady shape and the disconnection glide velocity goes to zero, $v_d = M_d [(\sigma+\tau_0)b -\gamma \partial_{xx} hH] =0$.
The GB shape profile (and the disconnection density distribution) may be implicitly determined from this expression.
The GB profile evolution is the same whether the GB is driven by an applied stress ($\tau_0$) or a jump in the chemical potential across the GB ($\Psi$) provided that $\Psi H= \tau_0 b$ \citep{zhang2017prl}.
However, when multiple disconnection modes are activated, the evolution depends on the nature of the driving forces.

We now examine the behavior based upon the multi-mode GB equation of motion.
For a GB in equilibrium, the glide velocities of all disconnections must vanish, i.e., $v_i=M_d [(\sigma+\tau)b_i + (\Psi-\gamma \partial_{xx} h)H_i]=0$ for all $i$.
Since the shear-coupling factors $\beta_i=-b_i/H_i$ of different disconnection modes will, in general, be distinct, this implies that there must be zero driving force associated with both the disconnection Burgers vector ($f_b=0$) and the GB step character ($f_h=0$).
Hence, in equilibrium, $\sigma=-\tau$ and $\Psi=\gamma \partial_{xx} h$.
Moreover, since the two ends are pinned at $x/L_0=0$ and 1, the equilibrium GB shape is a parabola (independent of the applied stress $\tau$), $h(x)=(\Psi/\gamma)(x^2-x)$ for the case of an isotropic  GB energy $\gamma$.
Note that in most materials $\gamma$  depends on  GB crystallography (e.g., the GB inclination $\partial_x h$ and grain misorientation); in this case the equilibrium shape will not be a parabola but may be found from $\Psi=\gamma(\partial_x h) \partial_{xx} h$.
We validate these predictions via simulations below.

\begin{figure}[!ht]
\centering
\subfloat[]{\includegraphics[width=0.33\textwidth]{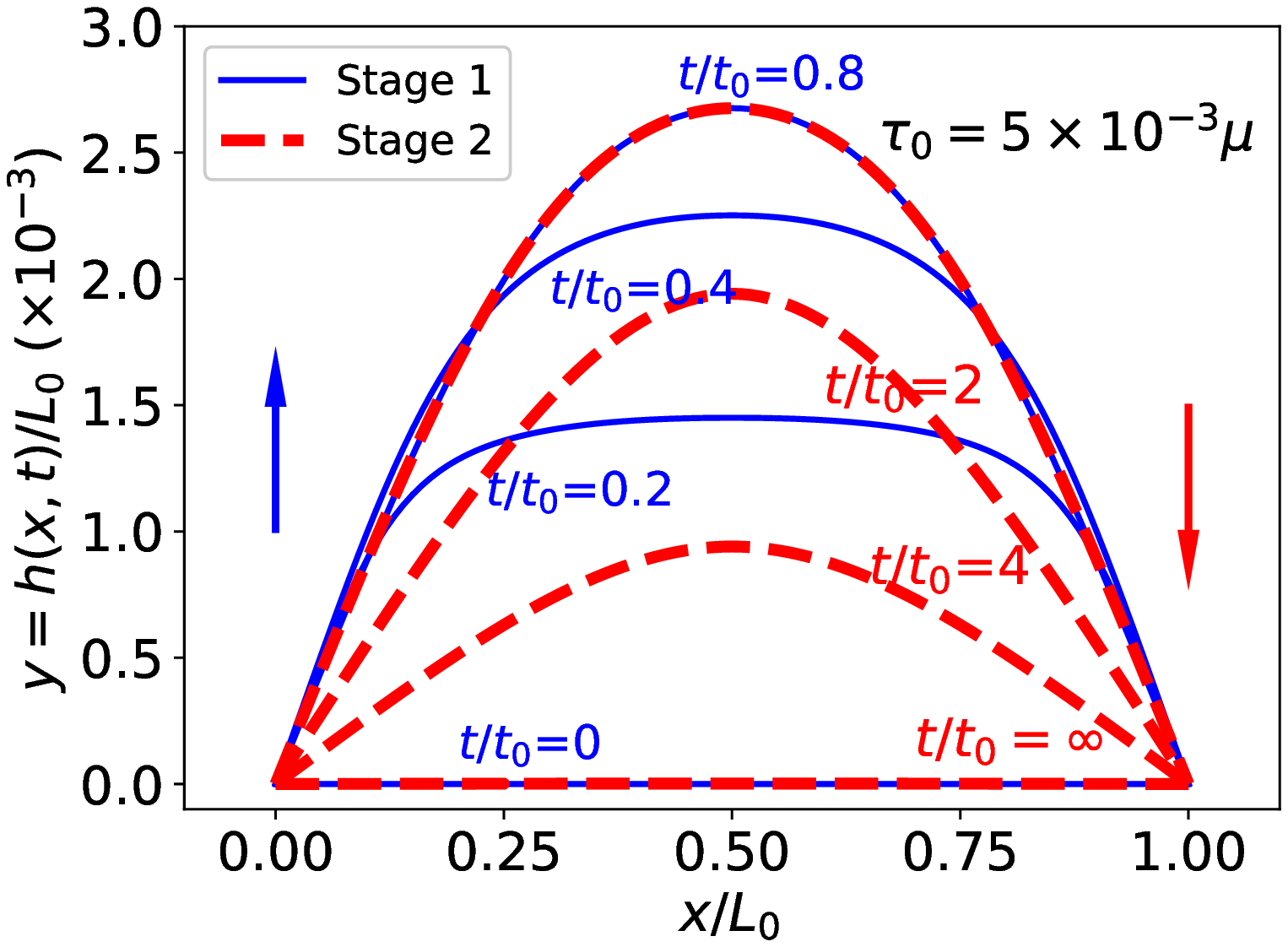}}
\subfloat[]{\includegraphics[width=0.33\textwidth]{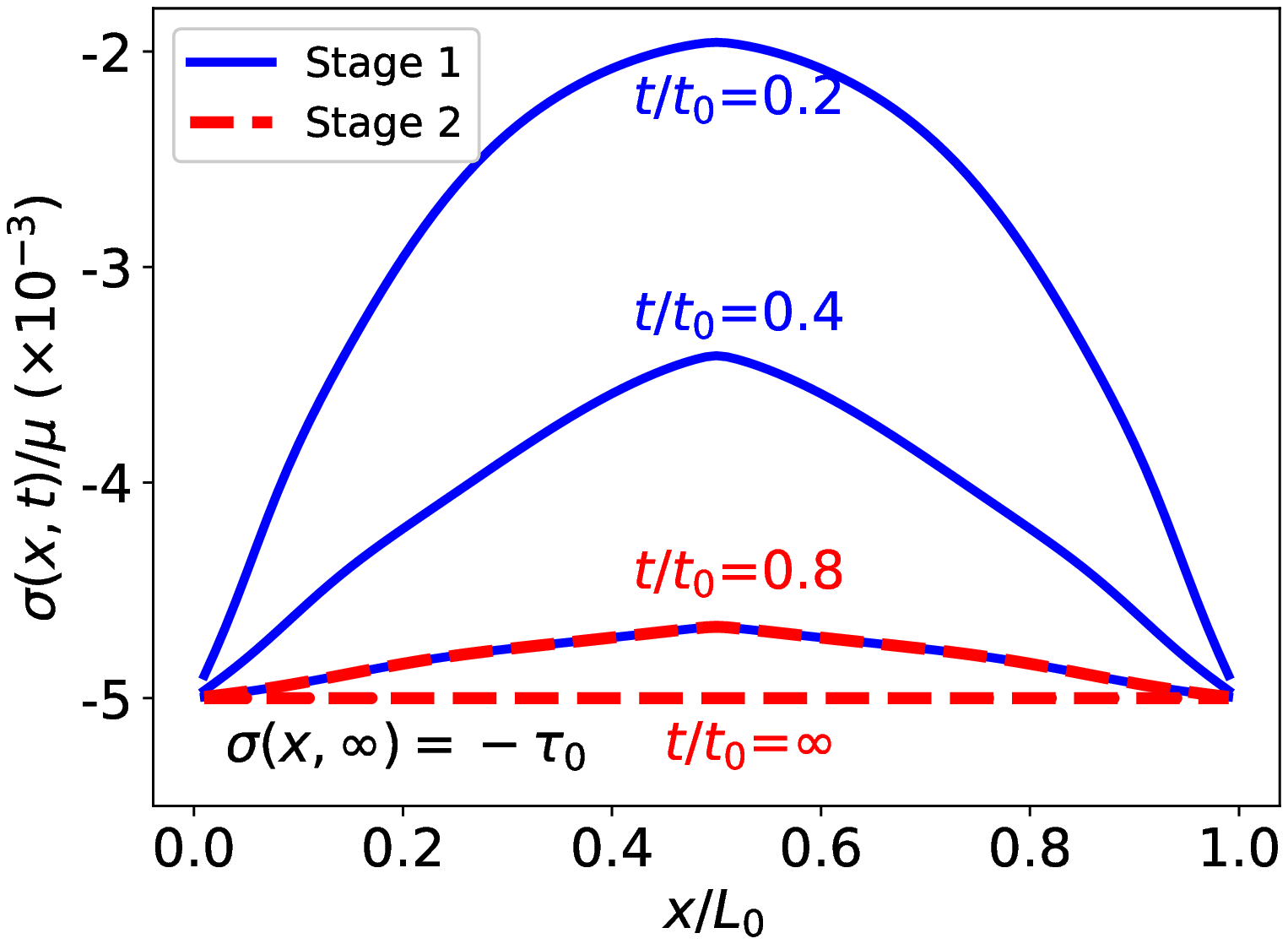}}
\subfloat[]{\includegraphics[width=0.33\textwidth]{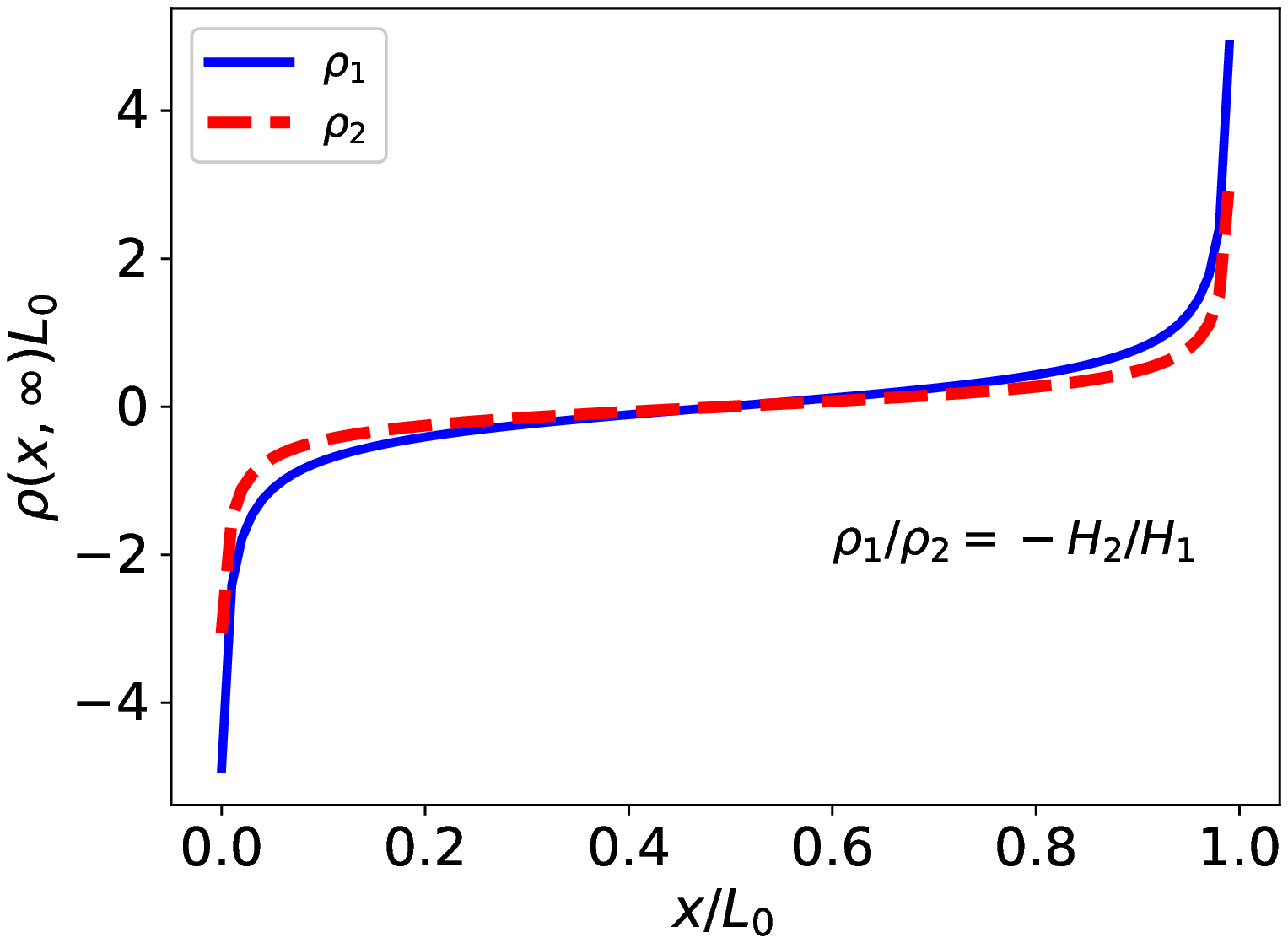}}
\caption{Two-mode migration of the $\Sigma13$ GB with pinned ends under an applied stress $\tau_0/\mu = 5\times10^{-3}$ at $T=800$~K.
(a) The evolution of this GB; the GB bows out at early time (stage 1, $t/t_0<0.8$) and then retracts back (stage 2, $t/t_0>0.8$).
Arrows indicate the GB migration directions during the two stages.
(b) The internal shear stress from all of the disconnections $\sigma(x,t)$.
At equilibrium, the internal stress balances  the external applied stress, i.e., $\sigma(x,\infty)=-\tau_0$.
(c) The disconnection density distribution at equilibrium ($\rho_1(x,\infty)H_1+\rho_2(x,\infty)H_2=0$).}
\label{fig:two_tau}
\end{figure}

We now consider the simple case in which only two disconnection modes with opposite shear-coupling factors (two-mode GB migration) are activated.
In particular, consider the stress-driven motion of a $\Sigma13$ GB in a face centered cubic material with the two lowest disconnection formation energy modes (and opposite shear coupling factors); i.e., ($a_0/\sqrt{26},-5a_0/2\sqrt{26}$) and ($a_0/\sqrt{26},4a_0/\sqrt{26}$), where $a_0$ is the cubic lattice parameter.
We observe two stages of evolution in Fig.~\ref{fig:two_tau}(a): in the first stage, the GB bows out quickly (as in the single-mode case, Fig.~\ref{fig:single_tau}) followed by a second stage in which the GB slowly flattens/retracts, eventually returning to its initially flat shape.
From the observation of the stresses $\sigma(x,t)$ from the disconnections along the GB  Fig.~\ref{fig:two_tau}(b), we find that during the first stage,  disconnections glide along the GB mainly driven by the external applied stress and rapidly reach a state such that the internal stress {\em nearly} offsets the external applied stress, i.e., $\sigma(t)\approx -\tau_0$.
In the second stage, the GB migration is mainly driven by the capillary force (i.e., the $\partial_{xx} h$ term).
Although the GB eventually returns to a flat shape, the equilibrium distributions of the disconnection densities are nonzero along the GB and satisfy the zero net step condition $\rho_1H_1+\rho_2H_2=0$ (as determined by $\partial_{xx} h=0$) while the net Burgers vector density  $\rho_1b_1+\rho_2b_2\neq 0$ (as determined by $\sigma=-\tau_0$), as illustrated in Fig.~\ref{fig:two_tau}(c).

\begin{figure}[!ht]
\centering
\subfloat[]{\includegraphics[width=0.33\textwidth]{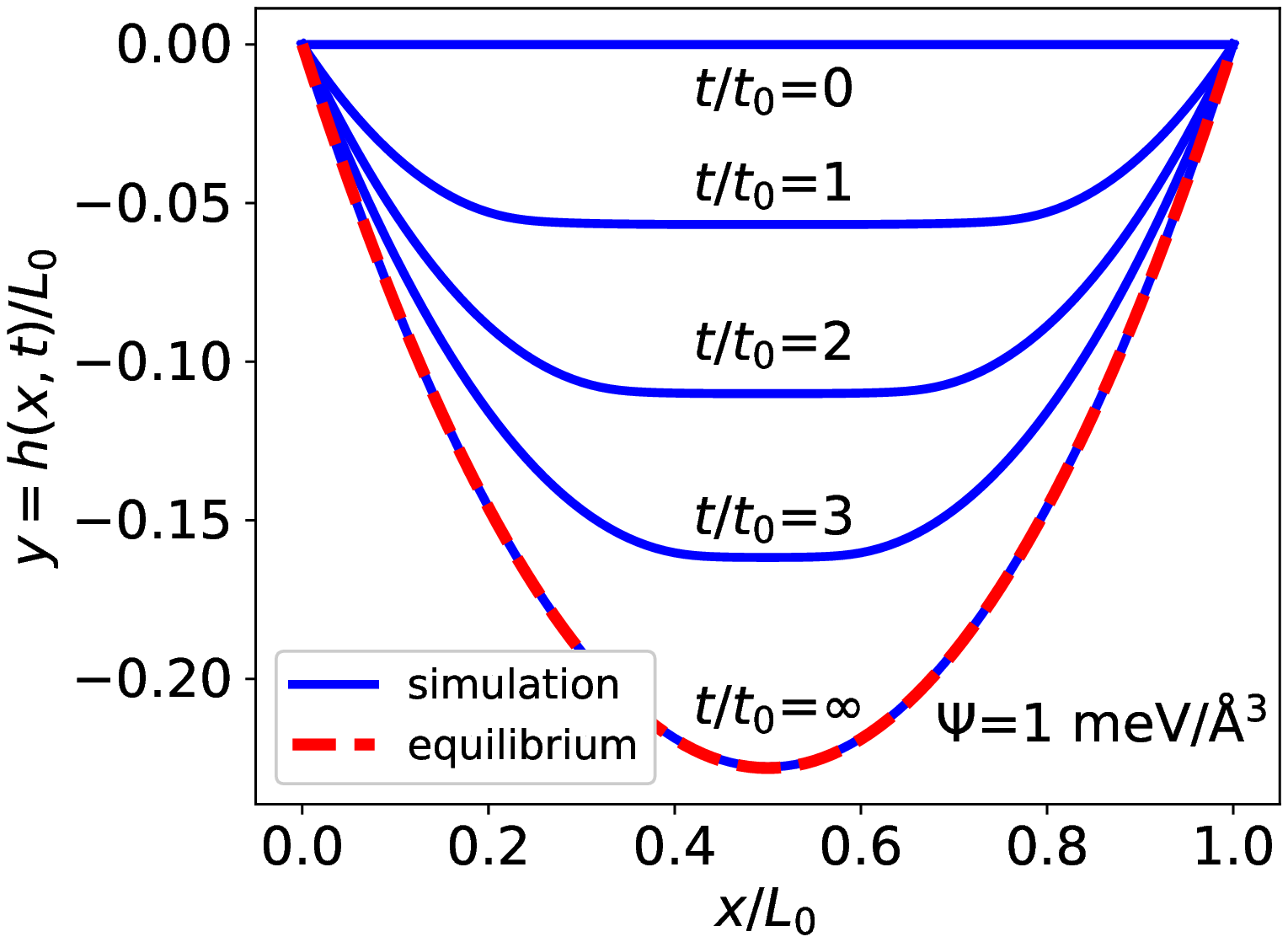}}
\subfloat[]{\includegraphics[width=0.33\textwidth]{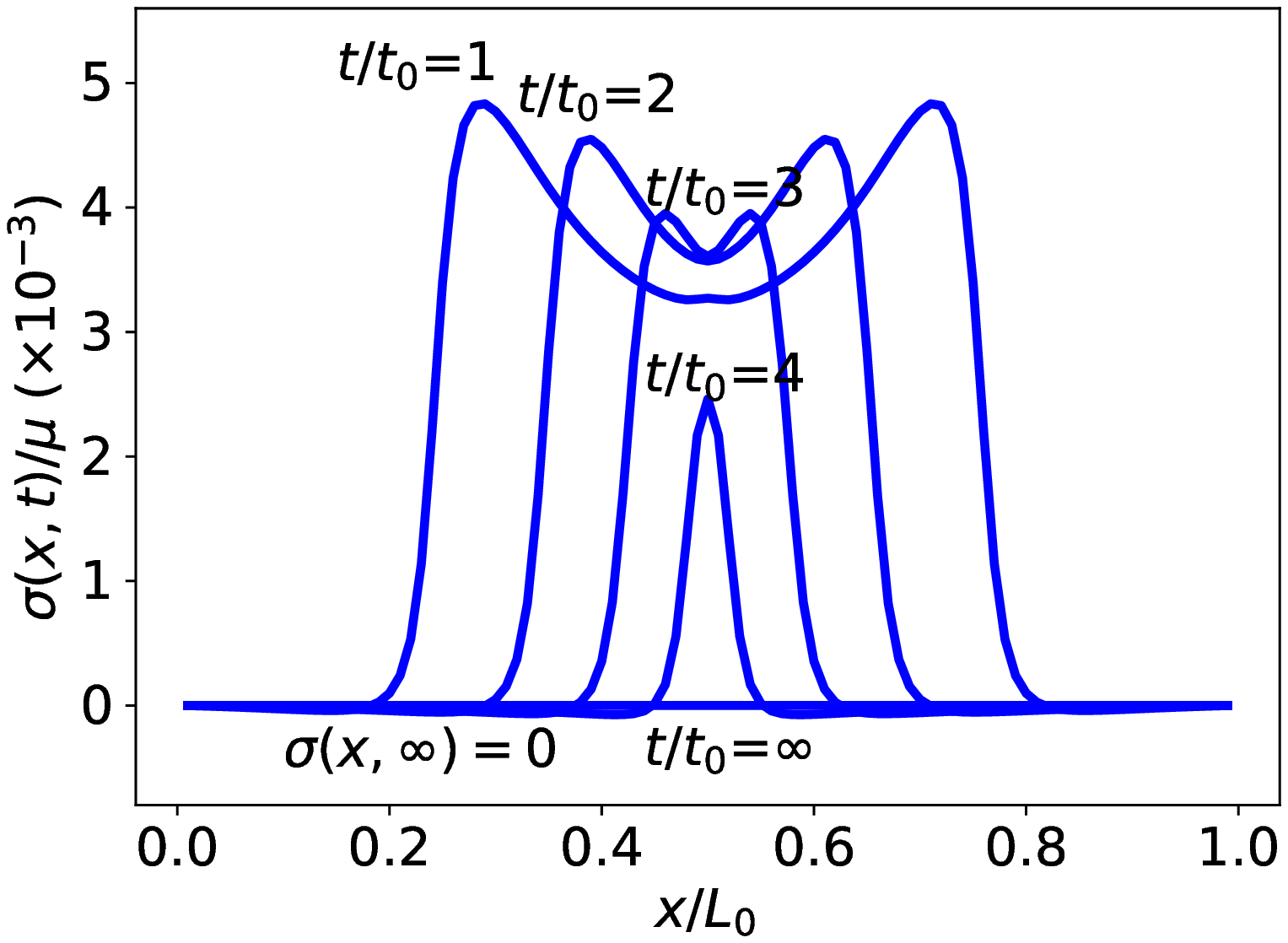}}
\subfloat[]{\includegraphics[width=0.33\textwidth]{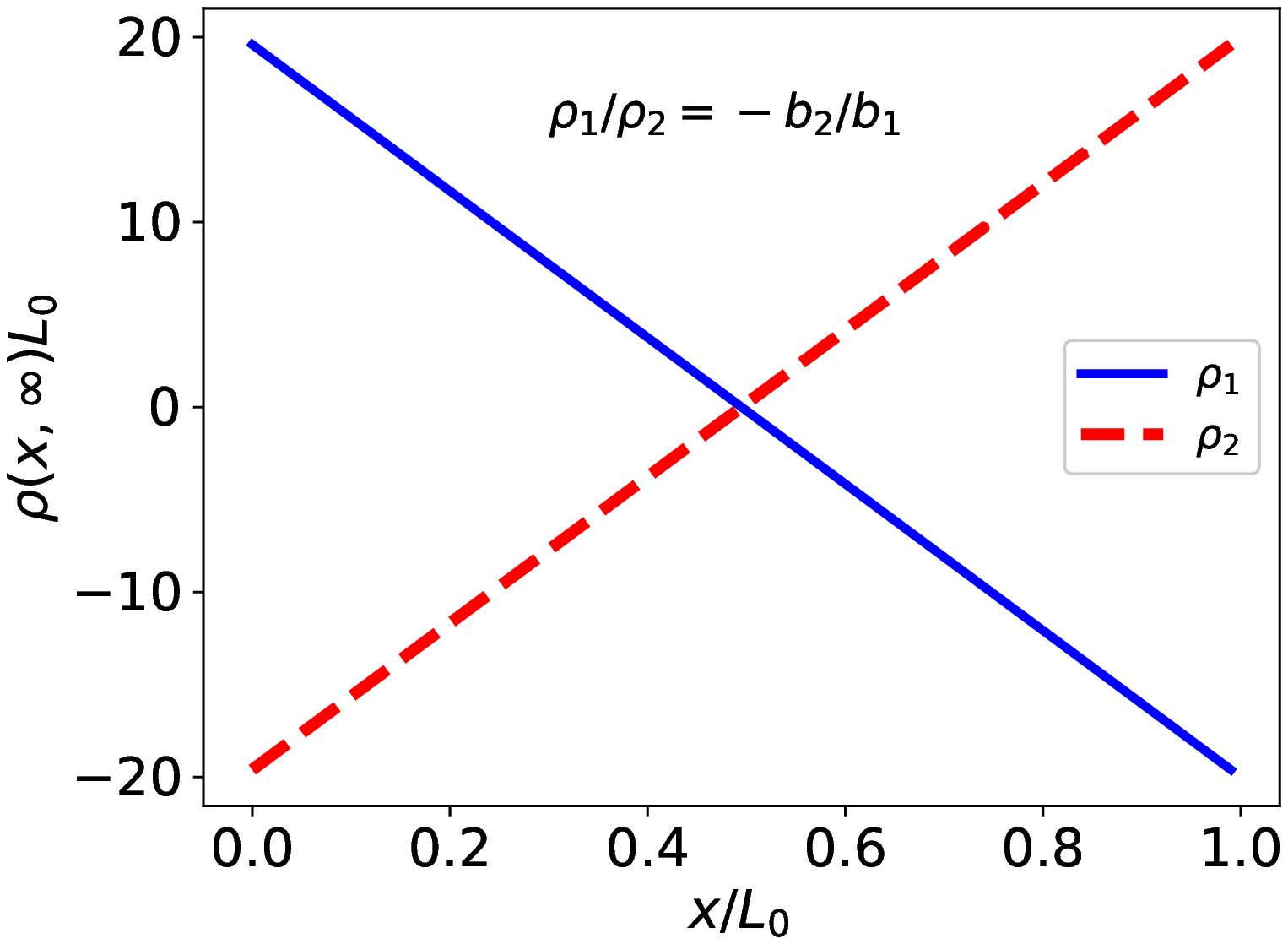}}
\caption{Two-mode migration of the $\Sigma13$ GB with pinned ends driven by a jump in chemical potential.
(a) The evolution of the $\Sigma$13 GB driven with $\Psi=1$ meV/\AA$^3$ at $T=800$~K.
(b) The internal shear stress associated with the disconnections $\sigma(x,t)$. At equilibrium, the internal stress vanishes, i.e., $\sigma(x,t=\infty)=0$.
(c) The equilibrium disconnection density distributions: $\rho_1(x,\infty)b_1+\rho_2(x,\infty)b_2=0$ implying that there is zero net Burgers vector along GB.
}\label{fig:two_Phi}
\end{figure}

Next, we drive the same  two disconnection mode GB with a jump in chemical potential (and zero applied stress).
As shown in Fig.~\ref{fig:two_Phi}(a), the GB bows out from its flat initial state,  attaining a stationary profile at late time that is  parabolic in agreement with our theoretical prediction.
In this  equilibrium state, the stress along the GB $\sigma=0$ (see Fig.~\ref{fig:two_Phi}(b)) and the capillarity driving force balances  the chemical potential jump ($\gamma \partial_{xx} h=\Psi $).
The corresponding equilibrium disconnection density satisfy the conditions $\rho_1b_1+\rho_2b_2=0$ and $\rho_1H_1+\rho_2H_2=\partial_x h=(\Psi/\gamma)(2x-1)$; see Fig.~\ref{fig:two_Phi}(c).

\begin{figure}[!ht]
\centering
\includegraphics[width=0.5\textwidth]{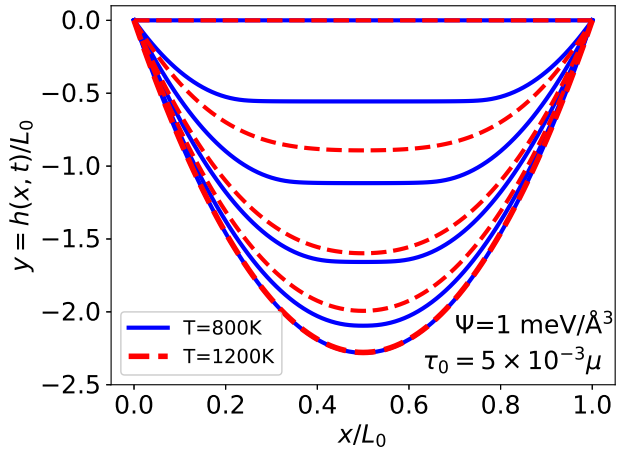}
\caption{Multi-mode  migration of a $\Sigma$13 GB with pinned ends at $T=800$~K and 1200~K under mixed driving forces; i.e., $\tau_0/\mu = 5\times10^{-3}$ and $\Psi=1$ meV/\AA$^3$.
Solid lines represent the GB evolution at $T=800$~K at $t/t_0=1,2,3,\infty$; dashed lines represent the GB evolution at $T=1200$~K at $t/t_0=0.1,0.2,0.3,\infty$.
}\label{fig:multi_mixed}
\end{figure}

As a general example of multi-mode GB migration, we examine the $\Sigma$13 GB with all disconnection modes; see Fig.~\ref{fig:dis_energy}.
Figure~\ref{fig:multi_mixed} shows the evolution of the GB under mixed driving forces, simultaneously including an applied stress and synthetic force at $T=800$~K and 1200~K.
The GB migrates to the same parabolic equilibrium shape at both temperatures, but with a much faster velocity at higher temperature.
The results indicate that the multi-mode GB migration is similar to the two-mode GB migration example in Figs.~\ref{fig:two_tau} and \ref{fig:two_Phi} (despite the difference in the equilibrium disconnection density distribution), but is very distinct from the single-mode GB migration results (Fig. \ref{fig:single_tau}).

While  all disconnection modes may play a role in GB migration and sliding, the importance of the different disconnection modes depend on their relative ease of nucleation (nucleation barrier, heterogeneous sites, interactions with lattice defects), which varies between GBs of different bicrystallography and atomic bonding, and the nature of the driving force.
This implies that different GBs will behave differently during the evolution of polycrystalline microstructures and these  differences will be a function of temperature.
The present results also demonstrate that grain size also matters; triple junction effects will be much more important in nanocrystalline materials where each grain boundary length (area) is small, as compared with large grained materials.

\section{SUMMARY AND DISCUSSION}
\label{sec:summary}
Discrete models of grain boundary migration based upon disconnection motion have previously been proposed (e.g., see \citep{han2018}) and have enjoyed considerable success.
However, if our goal is to describe GB migration in a microstructure, tracking discrete disconnections throughout the entire microstructure is not practical.
The main goal of this study has been the development of a continuum model of GB dynamics that respects the underlying microscopic mechanisms of GB migration (i.e., disconnection motion) without the excessive computational demands of tracking every disconnection.

We presented such a continuum model for  grain boundary dynamics (including migration and shear) based upon multiple disconnection modes which is a generalization of an earlier single disconnection mode model \citep{zhang2017prl}.
In this model,  GBs migrate via the thermally-activated nucleation and motion of  disconnections of different modes along the GBs.
The key to the temperature-dependence of GB dynamics is the competition and/or synergies of the different disconnection modes.
Grain boundary migration is, in general, coupled to shear deformation through disconnection motion, although different disconnection modes may conspire to decouple these.
We implement our model in continuum numerical simulations for GB dynamics in  single and multi-mode disconnection limits  in both a bicrystal (under two types of boundary conditions) and for a finite-length GB with pinned ends.
The simulation results demonstrate that the selection of and competition between different disconnection modes lead to complex and diverse shear-coupled GB migration behavior (and hence  grain growth) in a polycrystalline material.
When the shear deformation is constrained by fixed-surface BCs, as a model for a GB in a polycrystal, shear-coupling  implies stress generation that can lead to GB stagnation at low temperature, where only a single disconnection mode is active.
This constraint and the concomitant stress generation may be accommodated by the cooperation of multiple disconnection modes at high temperature such that GB motion can occur  without shear-coupling.
The temperature at which this transition in behavior occurs varies between GBs, depending on their relative disconnection nucleation barrier spectrum.


Our simulations provide clear demonstrations of the importance of including a crystallography-respecting, microscopic  model for microstructure evolution and the intrinsic coupling between stress, capillarity, and microstructure geometry in microstructure evolution.
Unlike in  conventional description of GB motion and grain growth, the GB mobility and shear coupling are not intrinsic properties of a GB but rather depend on the properties of their disconnections as well as how the GB is driven, microstructure geometry, and boundary conditions \citep{chen2019shear}.

The relative importance of different disconnection modes should, in general, depend on the relative ease of their nucleation and motion; these are functions of temperature and atomic-scale structure of the GBs.
The present model only considers GB motion controlled by disconnection nucleation; this implicitly assumes that disconnection motion is fast compared with disconnection migration (an assumption that is not always justified \citep{Combe2016kink}).
Further, the model for disconnection nucleation \citep{han2018} is based upon a model which may be oversimplified.
This implies that the prediction of GB dynamics based upon the disconnection model requires more detailed and accurate models for disconnection nucleation and migration barriers that provide more rigorous descriptions of GB structure, disconnection core structure and bonding.
Such input may be obtained from atomistic simulations that provide these barriers (e.g., using a transition barrier finding approach such as the nudged elastic band method  \citep{Combe2016kink}) and, ideally, incorporate accurate descriptions of atomic interactions (based upon first-principles methods).

Modeling the evolution of a polycrystalline microstructure requires a description of the spatial distribution of triple junctions, the topology of the microstructure, and grain size, as well as a description of how TJs move.
In this paper, we considered only stationary (immobile) TJs, that serve as fixed GB termini.
The disconnection description of GB dynamics has the potential to describe many of the complexities of microstructure evolution, including TJ dynamics \citep{han2018,zhang2017prl}.
Just as disconnection dynamics provides a robust approach for describing GB dynamics, we expect that it can also be extended to understand TJ motion, as recently proposed by \cite{thomas2019TJ}.

While a disconnection model for shear-coupled grain boundary migration is a promising approach for describing microstructure evolution, a number of important opportunities remain.
Although the present disconnection model of GB migration implicitly links microstructure evolution and mechanics, a more complete integration would couple grain boundary migration, grain boundary sliding,  grain rotation and stress generation.
This, together with better descriptions of disconnection dynamics at triple junctions and the dynamics of GBs of arbitrary inclination/shape, can be used to develop predictive models for microstructure evolution. Further, since the Hall-Petch effect (grain size strengthening)  is based on GBs  blocking  lattice dislocations, a more complete understanding of how GBs block, absorb, and transmit dislocation is important for describing plasticity in polycrystals.
Such effects are controlled, in part, by disconnection activity in the GB.
An extension of the present work to develop a crystallography-sensitive continuum model could be used to augment crystal plasticity descriptions of deformation to make such models microstructure-sensitive.
While GBs can absorb point defects (e.g., in radiation damage scenarios), some GBs readily absorb point defects while others do not.
This implies a connection between GB dynamics and point defect absorption efficiency.
A continuum disconnection-based model could couple GB adsorption efficiency, GB crystallography, and microstructure evolution.
This short list suggests that the coupling of microstructure evolution, defect dynamics within grains, and deformation within a disconnection-based, crystallography-respecting continuum description is a promising direction for both materials science and  the mechanics of materials.

\section*{Acknowledgement}
The research contributions of D.J.S and J.H were sponsored by the Army Research Office and were accomplished under Grant Number W911NF-19-1-0263. The views and conclusions contained in this document are those of the authors and should not be interpreted as representing the official policies, either expressed or implied, of the Army Research Office or the U.S. Government. The U.S. Government is authorized to reproduce and distribute reprints for Government purposes notwithstanding any copyright notation herein. Y.X. acknowledges support from the Hong Kong Research Grants Council General Research Fund 16302818.

\appendix
\section{Elasticity solutions}
\label{appendix:elst}
The stress and displacement fields for a period array of dislocations 
(periodicity $2w$) are given in \cite{hirth1982theory} and \cite{needleman1995discrete}
\begin{equation}\label{eq:period_soln2}
\begin{aligned}
&\sigma_{11} = -\frac{\mu b}{4w(1-\nu)}\frac{1}{\cosh(\pi \eta)-\cos(\pi\xi)}\Big[2\sinh(\pi\eta)+\pi\eta\frac{1-\cos(\pi\xi)\cosh(\pi\eta)}{\cosh(\pi \eta)-\cos(\pi\xi)}\Big],\\[4pt]
&\sigma_{22} = \frac{\mu b}{4w(1-\nu)}\Big[\pi\eta\frac{1-\cos(\pi\xi)\cosh(\pi\eta)}{\big(\cosh(\pi \eta)-\cos(\pi\xi)\big)^2}\Big], \\[4pt]
&\sigma_{12} = \frac{\mu b}{4w(1-\nu)}\frac{\sin(\pi\xi)}{\cosh(\pi \eta)-\cos(\pi\xi)}\Big[1-\pi\eta\frac{\sinh(\pi\eta)}{\cosh(\pi \eta)-\cos(\pi\xi)}\Big], \\[4pt]
&u_1(x,y) = 
\frac{b}{2\pi(1-\nu)}\Big[\frac{1}{4}\frac{\pi\eta\sin(\pi\xi)}{\cosh(\pi\eta)-\cos(\pi\xi)}-(1-\nu)\tan^{-1}\Big(\frac{\tan(\pi\xi/2)}{\tanh(\pi\eta/2)}\Big)\Big] ,\\[4pt]
&u_2(x,y) = \frac{b}{2\pi(1-\nu)}\Big[\frac{1}{4}\frac{\pi\eta\sinh(\pi\eta)}{\cosh(\pi\eta)-\cos(\pi\xi)}-\frac{1}{4}(1-2\nu)\ln\Big(\cosh(\pi\eta)-\cos(\pi\xi)\Big)\Big],
\end{aligned}
\end{equation}
where
\begin{equation}
\xi = \frac{x-x_0}{w}, \quad \eta=\frac{y-y_0}{w}.
\end{equation} \par

A Fourier series description of the Airy stress function for  free-surface boundary condition is as follows:
\begin{equation} \label{eq:soln_free2}
\begin{aligned}
&\varphi(x,y) = \sum_k \hat{\varphi} e^{ikx}, \\[4pt]
&\hat{\varphi}(k,y) = (A_1+B_1y)\cosh(ky) + (A_2+B_2y)\sinh(ky), \\[4pt]
& A_1 = -\frac{1}{\cosh(kL_y)}\Big(\frac{\hat{g}_1+\hat{g}_2}{2k^2}+\frac{L_y}{k}\Psi_2\Big),\quad B_1 = \frac{1}{k\cosh(kL_y)}\Psi_1, \\[4pt]
& A_2 = -\frac{1}{\sinh(kL_y)}\Big(\frac{\hat{g}_1-\hat{g}_2}{2k^2}+\frac{L_y}{k}\Psi_1\Big),\quad B_2 = \frac{1}{k\cosh(kL_y)}\Psi_2, \\[4pt]
&\Psi_1 = \frac{i\tanh(kL_y)(\hat{f}_1+\hat{f}_2)+(\hat{g}_1-\hat{g}_2)}{2kL_y\big(\tanh^2(kL_y)-1\big)+2\tanh(kL_y)},\\[4pt]
&\Psi_2 = \frac{i(\hat{f}_1-\hat{f}_2)+\tanh(kL_y)(\hat{g}_1+\hat{g}_2)}{2kL_y\big(1-\tanh^2(kL_y)\big)+2\tanh(kL_y)},
\end{aligned}
\end{equation}
where $\hat{f}_i$ and $\hat{g}_i$ are  Fourier coefficients of $f_i$ and $g_i$ for $i=1,2$, respectively.
We can easily apply this stress function to obtain the stress field in the usual manner as
\begin{equation}\label{eq:airy_fun}
\sigma_{11}=\frac{\partial^2\varphi}{\partial y^2},\quad
\sigma_{12}=-\frac{\partial^2\varphi}{\partial x\partial y},\quad
\sigma_{22}=\frac{\partial^2\varphi}{\partial x^2}.
\end{equation}\par

The Fourier coefficients of the displacement field for fixed-surface boundary condition
\begin{equation}\label{eq:soln_fix2}
\begin{aligned}
\hat{u}_1(k,y) = (A_1+B_1y)\cosh(ky) + (C_1+D_1y)\sinh(ky), \\
\hat{u}_2(k,y) = (A_2+B_2y)\cosh(ky) + (C_2+D_2y)\sinh(ky),
\end{aligned}
\end{equation}
where the coefficients are
\begin{equation}\label{eq:soln_fix3}
\begin{aligned}
& A_1 = i\Big[C_2+\frac{3-4\nu}{k}B_2\Big], \quad B_1 = iD_2, \\[4pt]
& C_1 = i\Big[A_2+\frac{3-4\nu}{k}D_2\Big], \quad D_1 = iB_2, \\[4pt]
& A_2 = \frac{\hat{g}_1+\hat{g}_2}{2\cosh(kL_y)}-L_y\tanh(kL_y)D_2, \quad B_2 = \frac{k}{\cosh(kL_y)}\Psi_1, \\[4pt]
& C_2 = \frac{\hat{g}_1-\hat{g}_2}{2\sinh(kL_y)}-L_y\coth(kL_y)B_2, \quad D_2 = \frac{-k}{\cosh(kL_y)}\Psi_2, \\[4pt]
&\Psi_1 = \frac{i\tanh(kL_y)(\hat{f}_1+\hat{f}_2)+(\hat{g}_1-\hat{g}_2)}{2kL_y\big(1-\tanh^2(kL_y)\big)-2(3-4\nu)\tanh(kL_y)} \text{sgn}(\eta),\\[4pt]
&\Psi_2 = \frac{i(\hat{f}_1-\hat{f}_2)+\tanh(kL_y)(\hat{g}_1+\hat{g}_2)}{2kL_y\big(1-\tanh^2(kL_y)\big)+2(3-4\nu)\tanh(kL_y)}.
\end{aligned}
\end{equation}

\bibliographystyle{elsarticle-harv}
\bibliography{myreference}

\begin{thebibliography}{68}
\expandafter\ifx\csname natexlab\endcsname\relax\def\natexlab#1{#1}\fi
\providecommand{\url}[1]{\texttt{#1}}
\providecommand{\href}[2]{#2}
\providecommand{\path}[1]{#1}
\providecommand{\DOIprefix}{doi:}
\providecommand{\ArXivprefix}{arXiv:}
\providecommand{\URLprefix}{URL: }
\providecommand{\Pubmedprefix}{pmid:}
\providecommand{\doi}[1]{\href{http://dx.doi.org/#1}{\path{#1}}}
\providecommand{\Pubmed}[1]{\href{pmid:#1}{\path{#1}}}
\providecommand{\bibinfo}[2]{#2}
\ifx\xfnm\relax \def\xfnm[#1]{\unskip,\space#1}\fi
\bibitem[{Ashby(1972)}]{Ashby1972498}
\bibinfo{author}{Ashby, M.}, \bibinfo{year}{1972}.
\newblock \bibinfo{title}{Boundary defects, and atomistic aspects of boundary
  sliding and diffusional creep}.
\newblock \bibinfo{journal}{Surface Science} \bibinfo{volume}{31},
  \bibinfo{pages}{498 -- 542}.
\newblock \URLprefix
  \url{http://www.sciencedirect.com/science/article/pii/0039602872902737},
  \DOIprefix\doi{https://doi.org/10.1016/0039-6028(72)90273-7}.
\bibitem[{Bainbridge et~al.(1954)Bainbridge, Choh and
  Edwards}]{Bainbridge1954shear}
\bibinfo{author}{Bainbridge, D.W.}, \bibinfo{author}{Choh, H.L.},
  \bibinfo{author}{Edwards, E.H.}, \bibinfo{year}{1954}.
\newblock \bibinfo{title}{Recent observations on the motion of small angle
  dislocation boundaries}.
\newblock \bibinfo{journal}{Acta Metallurgica} \bibinfo{volume}{2},
  \bibinfo{pages}{322 -- 333}.
\newblock \URLprefix
  \url{http://www.sciencedirect.com/science/article/pii/0001616054901753},
  \DOIprefix\doi{https://doi.org/10.1016/0001-6160(54)90175-3}.
\bibitem[{Bollmann(1970)}]{bollmann1970crystal}
\bibinfo{author}{Bollmann, W.}, \bibinfo{year}{1970}.
\newblock \bibinfo{title}{Crystal Defects and Crystalline Interfaces}.
\newblock \bibinfo{publisher}{Springer Berlin Heidelberg}.
\newblock \URLprefix \url{https://books.google.com.hk/books?id=i4TZRgAACAAJ}.
\bibitem[{Cahn et~al.(2006)Cahn, Mishin and Suzuki}]{cahn2006coupling}
\bibinfo{author}{Cahn, J.W.}, \bibinfo{author}{Mishin, Y.},
  \bibinfo{author}{Suzuki, A.}, \bibinfo{year}{2006}.
\newblock \bibinfo{title}{Coupling grain boundary motion to shear deformation}.
\newblock \bibinfo{journal}{Acta Materialia} \bibinfo{volume}{54},
  \bibinfo{pages}{4953 -- 4975}.
\newblock \URLprefix
  \url{http://www.sciencedirect.com/science/article/pii/S1359645406005313},
  \DOIprefix\doi{https://doi.org/10.1016/j.actamat.2006.08.004}.
\bibitem[{Cahn and Taylor(2004)}]{cahn2004unified}
\bibinfo{author}{Cahn, J.W.}, \bibinfo{author}{Taylor, J.E.},
  \bibinfo{year}{2004}.
\newblock \bibinfo{title}{A unified approach to motion of grain boundaries,
  relative tangential translation along grain boundaries, and grain rotation}.
\newblock \bibinfo{journal}{Acta Materialia} \bibinfo{volume}{52},
  \bibinfo{pages}{4887 -- 4898}.
\newblock \URLprefix
  \url{http://www.sciencedirect.com/science/article/pii/S1359645404003945},
  \DOIprefix\doi{https://doi.org/10.1016/j.actamat.2004.02.048}.
\bibitem[{Chen et~al.(2019)Chen, Han, Thomas and Srolovitz}]{chen2019shear}
\bibinfo{author}{Chen, K.}, \bibinfo{author}{Han, J.}, \bibinfo{author}{Thomas,
  S.L.}, \bibinfo{author}{Srolovitz, D.J.}, \bibinfo{year}{2019}.
\newblock \bibinfo{title}{Grain boundary shear coupling is not a grain boundary
  property}.
\newblock \bibinfo{journal}{Acta Materialia} \bibinfo{volume}{167},
  \bibinfo{pages}{241 -- 247}.
\newblock \URLprefix
  \url{http://www.sciencedirect.com/science/article/pii/S1359645419300552},
  \DOIprefix\doi{https://doi.org/10.1016/j.actamat.2019.01.040}.
\bibitem[{Chen and Kalonji(1992)}]{chenlq1992MD}
\bibinfo{author}{Chen, L.Q.}, \bibinfo{author}{Kalonji, G.},
  \bibinfo{year}{1992}.
\newblock \bibinfo{title}{Finite temperature structure and properties of ∑ =
  5 (310) tilt grain boundaries in nacl a molecular dynamics study}.
\newblock \bibinfo{journal}{Philosophical Magazine A} \bibinfo{volume}{66},
  \bibinfo{pages}{11--26}.
\newblock \URLprefix \url{https://doi.org/10.1080/01418619208201510},
  \DOIprefix\doi{10.1080/01418619208201510},
  \href{http://arxiv.org/abs/https://doi.org/10.1080/01418619208201510}{{\tt
  arXiv:https://doi.org/10.1080/01418619208201510}}.
\bibitem[{Chen and Yang(1994)}]{chenlq1994simulation}
\bibinfo{author}{Chen, L.Q.}, \bibinfo{author}{Yang, W.}, \bibinfo{year}{1994}.
\newblock \bibinfo{title}{Computer simulation of the domain dynamics of a
  quenched system with a large number of nonconserved order parameters: The
  grain-growth kinetics}.
\newblock \bibinfo{journal}{Physical Review B} \bibinfo{volume}{50},
  \bibinfo{pages}{15752--15756}.
\newblock \URLprefix \url{https://link.aps.org/doi/10.1103/PhysRevB.50.15752},
  \DOIprefix\doi{10.1103/PhysRevB.50.15752}.
\bibitem[{Combe et~al.(2016)Combe, Mompiou and Legros}]{Combe2016kink}
\bibinfo{author}{Combe, N.}, \bibinfo{author}{Mompiou, F.},
  \bibinfo{author}{Legros, M.}, \bibinfo{year}{2016}.
\newblock \bibinfo{title}{Disconnections kinks and competing modes in
  shear-coupled grain boundary migration}.
\newblock \bibinfo{journal}{Phys. Rev. B} \bibinfo{volume}{93},
  \bibinfo{pages}{024109}.
\newblock \URLprefix \url{https://link.aps.org/doi/10.1103/PhysRevB.93.024109},
  \DOIprefix\doi{10.1103/PhysRevB.93.024109}.
\bibitem[{{DeCost} and {Holm}(2017)}]{decost1017abnormal}
\bibinfo{author}{{DeCost}, B.L.}, \bibinfo{author}{{Holm}, E.A.},
  \bibinfo{year}{2017}.
\newblock \bibinfo{title}{{Phenomenology of Abnormal Grain Growth in Systems
  with Nonuniform Grain Boundary Mobility}}.
\newblock \bibinfo{journal}{Metallurgical and Materials Transactions A}
  \bibinfo{volume}{48}, \bibinfo{pages}{2771--2780}.
\newblock \DOIprefix\doi{10.1007/s11661-016-3673-6}.
\bibitem[{Elsey et~al.(2009)Elsey, Esedo\={g}lu and
  Smereka}]{elsey2009diffusion}
\bibinfo{author}{Elsey, M.}, \bibinfo{author}{Esedo\={g}lu, S.},
  \bibinfo{author}{Smereka, P.}, \bibinfo{year}{2009}.
\newblock \bibinfo{title}{Diffusion generated motion for grain growth in two
  and three dimensions}.
\newblock \bibinfo{journal}{Journal of Computational Physics}
  \bibinfo{volume}{228}, \bibinfo{pages}{8015 -- 8033}.
\newblock \URLprefix
  \url{http://www.sciencedirect.com/science/article/pii/S0021999109004082},
  \DOIprefix\doi{https://doi.org/10.1016/j.jcp.2009.07.020}.
\bibitem[{Esedo\={g}lu(2016)}]{esedoglu2016grain}
\bibinfo{author}{Esedo\={g}lu, S.}, \bibinfo{year}{2016}.
\newblock \bibinfo{title}{Grain size distribution under simultaneous grain
  boundary migration and grain rotation in two dimensions}.
\newblock \bibinfo{journal}{Computational Materials Science}
  \bibinfo{volume}{121}, \bibinfo{pages}{209--216}.
\bibitem[{Van~der Giessen and Needleman(1995)}]{needleman1995discrete}
\bibinfo{author}{Van~der Giessen, E.}, \bibinfo{author}{Needleman, A.},
  \bibinfo{year}{1995}.
\newblock \bibinfo{title}{Discrete dislocation plasticity: a simple planar
  model}.
\newblock \bibinfo{journal}{Modelling and Simulation in Materials Science and
  Engineering} \bibinfo{volume}{3}, \bibinfo{pages}{689}.
\newblock \URLprefix \url{http://stacks.iop.org/0965-0393/3/i=5/a=008}.
\bibitem[{Gottstein et~al.(2001)Gottstein, Molodov, Shvindlerman, Srolovitz and
  Winning}]{gottstein2001grain}
\bibinfo{author}{Gottstein, G.}, \bibinfo{author}{Molodov, D.},
  \bibinfo{author}{Shvindlerman, L.}, \bibinfo{author}{Srolovitz, D.},
  \bibinfo{author}{Winning, M.}, \bibinfo{year}{2001}.
\newblock \bibinfo{title}{Grain boundary migration: misorientation dependence}.
\newblock \bibinfo{journal}{Current Opinion in Solid State and Materials
  Science} \bibinfo{volume}{5}, \bibinfo{pages}{9 -- 14}.
\newblock \URLprefix
  \url{http://www.sciencedirect.com/science/article/pii/S1359028600000309},
  \DOIprefix\doi{https://doi.org/10.1016/S1359-0286(00)00030-9}.
\bibitem[{Hamilton and Foiles(2002)}]{Hamilton2002prb}
\bibinfo{author}{Hamilton, J.C.}, \bibinfo{author}{Foiles, S.M.},
  \bibinfo{year}{2002}.
\newblock \bibinfo{title}{First-principles calculations of grain boundary
  theoretical shear strength using transition state finding to determine
  generalized gamma surface cross sections}.
\newblock \bibinfo{journal}{Physical Review B} \bibinfo{volume}{65},
  \bibinfo{pages}{064104}.
\newblock \URLprefix \url{https://link.aps.org/doi/10.1103/PhysRevB.65.064104},
  \DOIprefix\doi{10.1103/PhysRevB.65.064104}.
\bibitem[{Han et~al.(2018)Han, Thomas and Srolovitz}]{han2018}
\bibinfo{author}{Han, J.}, \bibinfo{author}{Thomas, S.L.},
  \bibinfo{author}{Srolovitz, D.J.}, \bibinfo{year}{2018}.
\newblock \bibinfo{title}{Grain-boundary kinetics: A unified approach}.
\newblock \bibinfo{journal}{Progress in Materials Science} ,
  \bibinfo{pages}{--}\URLprefix
  \url{https://www.sciencedirect.com/science/article/pii/S0079642518300641},
  \DOIprefix\doi{https://doi.org/10.1016/j.pmatsci.2018.05.004}.
\bibitem[{Han et~al.(2017)Han, Vitek and Srolovitz}]{han2017redux}
\bibinfo{author}{Han, J.}, \bibinfo{author}{Vitek, V.},
  \bibinfo{author}{Srolovitz, D.J.}, \bibinfo{year}{2017}.
\newblock \bibinfo{title}{The grain-boundary structural unit model redux}.
\newblock \bibinfo{journal}{Acta Materialia} \bibinfo{volume}{133},
  \bibinfo{pages}{186 -- 199}.
\newblock \URLprefix
  \url{http://www.sciencedirect.com/science/article/pii/S1359645417303798},
  \DOIprefix\doi{https://doi.org/10.1016/j.actamat.2017.05.002}.
\bibitem[{Harris et~al.(1998)Harris, Singh and King}]{harris1998rotation}
\bibinfo{author}{Harris, K.}, \bibinfo{author}{Singh, V.},
  \bibinfo{author}{King, A.}, \bibinfo{year}{1998}.
\newblock \bibinfo{title}{Grain rotation in thin films of gold}.
\newblock \bibinfo{journal}{Acta Materialia} \bibinfo{volume}{46},
  \bibinfo{pages}{2623 -- 2633}.
\newblock \URLprefix
  \url{http://www.sciencedirect.com/science/article/pii/S1359645497004679},
  \DOIprefix\doi{https://doi.org/10.1016/S1359-6454(97)00467-9}.
\bibitem[{Hillert(1965)}]{hillert1965on}
\bibinfo{author}{Hillert, M.}, \bibinfo{year}{1965}.
\newblock \bibinfo{title}{On the theory of normal and abnormal grain growth}.
\newblock \bibinfo{journal}{Acta Metallurgica} \bibinfo{volume}{13},
  \bibinfo{pages}{227 -- 238}.
\newblock \URLprefix
  \url{http://www.sciencedirect.com/science/article/pii/0001616065902002},
  \DOIprefix\doi{https://doi.org/10.1016/0001-6160(65)90200-2}.
\bibitem[{Hirth and Pond(1996)}]{hirth1996disconnection}
\bibinfo{author}{Hirth, J.}, \bibinfo{author}{Pond, R.}, \bibinfo{year}{1996}.
\newblock \bibinfo{title}{Steps, dislocations and disconnections as interface
  defects relating to structure and phase transformations}.
\newblock \bibinfo{journal}{Acta Materialia} \bibinfo{volume}{44},
  \bibinfo{pages}{4749 -- 4763}.
\newblock \URLprefix
  \url{http://www.sciencedirect.com/science/article/pii/S1359645496001322},
  \DOIprefix\doi{https://doi.org/10.1016/S1359-6454(96)00132-2}.
\bibitem[{Hirth and Lothe(1982)}]{hirth1982theory}
\bibinfo{author}{Hirth, J.P.}, \bibinfo{author}{Lothe, J.},
  \bibinfo{year}{1982}.
\newblock \bibinfo{title}{Theory of dislocations}.
\newblock \bibinfo{edition}{2nd ed} ed., \bibinfo{publisher}{New York: Wiley}.
\bibitem[{Homer et~al.(2013)Homer, Foiles, Holm and Olmsted}]{Homer2013MD}
\bibinfo{author}{Homer, E.R.}, \bibinfo{author}{Foiles, S.M.},
  \bibinfo{author}{Holm, E.A.}, \bibinfo{author}{Olmsted, D.L.},
  \bibinfo{year}{2013}.
\newblock \bibinfo{title}{Phenomenology of shear-coupled grain boundary motion
  in symmetric tilt and general grain boundaries}.
\newblock \bibinfo{journal}{Acta Materialia} \bibinfo{volume}{61},
  \bibinfo{pages}{1048 -- 1060}.
\newblock \URLprefix
  \url{http://www.sciencedirect.com/science/article/pii/S1359645412007306},
  \DOIprefix\doi{https://doi.org/10.1016/j.actamat.2012.10.005}.
\bibitem[{Janssens et~al.(2006)Janssens, Olmsted, Holm, Foiles, Plimpton and
  Derlet}]{janssens2006computing}
\bibinfo{author}{Janssens, K.G.}, \bibinfo{author}{Olmsted, D.},
  \bibinfo{author}{Holm, E.A.}, \bibinfo{author}{Foiles, S.M.},
  \bibinfo{author}{Plimpton, S.J.}, \bibinfo{author}{Derlet, P.M.},
  \bibinfo{year}{2006}.
\newblock \bibinfo{title}{Computing the mobility of grain boundaries}.
\newblock \bibinfo{journal}{Nature materials} \bibinfo{volume}{5},
  \bibinfo{pages}{124}.
\bibitem[{Jin et~al.(2004)Jin, Minor, Stach and Morris}]{JIN2004direct}
\bibinfo{author}{Jin, M.}, \bibinfo{author}{Minor, A.}, \bibinfo{author}{Stach,
  E.}, \bibinfo{author}{Morris, J.}, \bibinfo{year}{2004}.
\newblock \bibinfo{title}{Direct observation of deformation-induced grain
  growth during the nanoindentation of ultrafine-grained al at room
  temperature}.
\newblock \bibinfo{journal}{Acta Materialia} \bibinfo{volume}{52},
  \bibinfo{pages}{5381 -- 5387}.
\newblock \URLprefix
  \url{http://www.sciencedirect.com/science/article/pii/S1359645404004653},
  \DOIprefix\doi{https://doi.org/10.1016/j.actamat.2004.07.044}.
\bibitem[{Kazaryan et~al.(2000)Kazaryan, Wang, Dregia and
  Patton}]{kazaryan2000phase}
\bibinfo{author}{Kazaryan, A.}, \bibinfo{author}{Wang, Y.},
  \bibinfo{author}{Dregia, S.A.}, \bibinfo{author}{Patton, B.R.},
  \bibinfo{year}{2000}.
\newblock \bibinfo{title}{Generalized phase-field model for computer simulation
  of grain growth in anisotropic systems}.
\newblock \bibinfo{journal}{Physical Review B} \bibinfo{volume}{61},
  \bibinfo{pages}{14275--14278}.
\newblock \URLprefix \url{https://link.aps.org/doi/10.1103/PhysRevB.61.14275},
  \DOIprefix\doi{10.1103/PhysRevB.61.14275}.
\bibitem[{Kinderlehrer et~al.(2006)Kinderlehrer, Livshits and
  Ta’asan}]{Kinderlehrer2006variation}
\bibinfo{author}{Kinderlehrer, D.}, \bibinfo{author}{Livshits, I.},
  \bibinfo{author}{Ta’asan, S.}, \bibinfo{year}{2006}.
\newblock \bibinfo{title}{A variational approach to modeling and simulation of
  grain growth}.
\newblock \bibinfo{journal}{SIAM Journal on Scientific Computing}
  \bibinfo{volume}{28}, \bibinfo{pages}{1694--1715}.
\newblock \URLprefix \url{https://doi.org/10.1137/030601971},
  \DOIprefix\doi{10.1137/030601971},
  \href{http://arxiv.org/abs/https://doi.org/10.1137/030601971}{{\tt
  arXiv:https://doi.org/10.1137/030601971}}.
\bibitem[{King and Smith(1980)}]{king1980}
\bibinfo{author}{King, A.H.}, \bibinfo{author}{Smith, D.A.},
  \bibinfo{year}{1980}.
\newblock \bibinfo{title}{{The effects on grain-boundary processes of the steps
  in the boundary plane associated with the cores of grain-boundary
  dislocations}}.
\newblock \bibinfo{journal}{Acta Crystallographica Section A}
  \bibinfo{volume}{36}, \bibinfo{pages}{335--343}.
\newblock \URLprefix \url{https://doi.org/10.1107/S0567739480000782},
  \DOIprefix\doi{10.1107/S0567739480000782}.
\bibitem[{Kobayashi et~al.(2000)Kobayashi, Warren and
  Carter}]{Kobayashi2000continuum}
\bibinfo{author}{Kobayashi, R.}, \bibinfo{author}{Warren, J.A.},
  \bibinfo{author}{Carter, W.C.}, \bibinfo{year}{2000}.
\newblock \bibinfo{title}{A continuum model of grain boundaries}.
\newblock \bibinfo{journal}{Physica D: Nonlinear Phenomena}
  \bibinfo{volume}{140}, \bibinfo{pages}{141 -- 150}.
\newblock \URLprefix
  \url{http://www.sciencedirect.com/science/article/pii/S0167278900000233},
  \DOIprefix\doi{https://doi.org/10.1016/S0167-2789(00)00023-3}.
\bibitem[{Lazar et~al.(2010)Lazar, MacPherson and
  Srolovitz}]{lazar2010accurate}
\bibinfo{author}{Lazar, E.A.}, \bibinfo{author}{MacPherson, R.D.},
  \bibinfo{author}{Srolovitz, D.J.}, \bibinfo{year}{2010}.
\newblock \bibinfo{title}{A more accurate two-dimensional grain growth
  algorithm}.
\newblock \bibinfo{journal}{Acta Materialia} \bibinfo{volume}{58},
  \bibinfo{pages}{364 -- 372}.
\newblock \URLprefix
  \url{http://www.sciencedirect.com/science/article/pii/S135964540900603X},
  \DOIprefix\doi{https://doi.org/10.1016/j.actamat.2009.09.008}.
\bibitem[{Legros et~al.(2008)Legros, Gianola and Hemker}]{legros2008abnormal}
\bibinfo{author}{Legros, M.}, \bibinfo{author}{Gianola, D.S.},
  \bibinfo{author}{Hemker, K.J.}, \bibinfo{year}{2008}.
\newblock \bibinfo{title}{In situ tem observations of fast grain-boundary
  motion in stressed nanocrystalline aluminum films}.
\newblock \bibinfo{journal}{Acta Materialia} \bibinfo{volume}{56},
  \bibinfo{pages}{3380 -- 3393}.
\newblock \URLprefix
  \url{http://www.sciencedirect.com/science/article/pii/S1359645408002152},
  \DOIprefix\doi{https://doi.org/10.1016/j.actamat.2008.03.032}.
\bibitem[{LeSar et~al.(1989)LeSar, Najafabadi and Srolovitz}]{LeSar1989finite}
\bibinfo{author}{LeSar, R.}, \bibinfo{author}{Najafabadi, R.},
  \bibinfo{author}{Srolovitz, D.J.}, \bibinfo{year}{1989}.
\newblock \bibinfo{title}{Finite-temperature defect properties from free-energy
  minimization}.
\newblock \bibinfo{journal}{Physical Review Letter} \bibinfo{volume}{63},
  \bibinfo{pages}{624--627}.
\newblock \URLprefix \url{https://link.aps.org/doi/10.1103/PhysRevLett.63.624},
  \DOIprefix\doi{10.1103/PhysRevLett.63.624}.
\bibitem[{Li et~al.(1953)Li, Edwards, Washburn and Parker}]{li1953stress}
\bibinfo{author}{Li, C.H.}, \bibinfo{author}{Edwards, E.},
  \bibinfo{author}{Washburn, J.}, \bibinfo{author}{Parker, E.},
  \bibinfo{year}{1953}.
\newblock \bibinfo{title}{Stress-induced movement of crystal boundaries}.
\newblock \bibinfo{journal}{Acta Metallurgica} \bibinfo{volume}{1},
  \bibinfo{pages}{223 -- 229}.
\newblock \URLprefix
  \url{http://www.sciencedirect.com/science/article/pii/0001616053900625},
  \DOIprefix\doi{https://doi.org/10.1016/0001-6160(53)90062-5}.
\bibitem[{Ma(2004)}]{Ma2004Science}
\bibinfo{author}{Ma, E.}, \bibinfo{year}{2004}.
\newblock \bibinfo{title}{Watching the nanograins roll}.
\newblock \bibinfo{journal}{Science} \bibinfo{volume}{305},
  \bibinfo{pages}{623--624}.
\newblock \URLprefix \url{http://science.sciencemag.org/content/305/5684/623},
  \DOIprefix\doi{10.1126/science.1101589},
  \href{http://arxiv.org/abs/http://science.sciencemag.org/content/305/5684/623.full.pdf}{{\tt
  arXiv:http://science.sciencemag.org/content/305/5684/623.full.pdf}}.
\bibitem[{MacPherson and Srolovitz(2007)}]{macpherson2007neumann}
\bibinfo{author}{MacPherson, R.D.}, \bibinfo{author}{Srolovitz, D.J.},
  \bibinfo{year}{2007}.
\newblock \bibinfo{title}{The von neumann relation generalized to coarsening of
  three-dimensional microstructures}.
\newblock \bibinfo{journal}{Nature} \bibinfo{volume}{446},
  \bibinfo{pages}{1053}.
\bibitem[{Mishin et~al.(2001)Mishin, Mehl, Papaconstantopoulos, Voter and
  Kress}]{Mishin2001Cu}
\bibinfo{author}{Mishin, Y.}, \bibinfo{author}{Mehl, M.J.},
  \bibinfo{author}{Papaconstantopoulos, D.A.}, \bibinfo{author}{Voter, A.F.},
  \bibinfo{author}{Kress, J.D.}, \bibinfo{year}{2001}.
\newblock \bibinfo{title}{Structural stability and lattice defects in copper:
  Ab initio, tight-binding, and embedded-atom calculations}.
\newblock \bibinfo{journal}{Phys. Rev. B} \bibinfo{volume}{63},
  \bibinfo{pages}{224106}.
\newblock \URLprefix \url{https://link.aps.org/doi/10.1103/PhysRevB.63.224106},
  \DOIprefix\doi{10.1103/PhysRevB.63.224106}.
\bibitem[{Molteni et~al.(1996)Molteni, Francis, Payne and
  Heine}]{Molteni1996prl}
\bibinfo{author}{Molteni, C.}, \bibinfo{author}{Francis, G.P.},
  \bibinfo{author}{Payne, M.C.}, \bibinfo{author}{Heine, V.},
  \bibinfo{year}{1996}.
\newblock \bibinfo{title}{First principles simulation of grain boundary
  sliding}.
\newblock \bibinfo{journal}{Physical Review Letter} \bibinfo{volume}{76},
  \bibinfo{pages}{1284--1287}.
\newblock \URLprefix
  \url{https://link.aps.org/doi/10.1103/PhysRevLett.76.1284},
  \DOIprefix\doi{10.1103/PhysRevLett.76.1284}.
\bibitem[{Molteni et~al.(1997)Molteni, Marzari, Payne and
  Heine}]{Molteni1997prl}
\bibinfo{author}{Molteni, C.}, \bibinfo{author}{Marzari, N.},
  \bibinfo{author}{Payne, M.C.}, \bibinfo{author}{Heine, V.},
  \bibinfo{year}{1997}.
\newblock \bibinfo{title}{Sliding mechanisms in aluminum grain boundaries}.
\newblock \bibinfo{journal}{Physical Review Letter} \bibinfo{volume}{79},
  \bibinfo{pages}{869--872}.
\newblock \URLprefix \url{https://link.aps.org/doi/10.1103/PhysRevLett.79.869},
  \DOIprefix\doi{10.1103/PhysRevLett.79.869}.
\bibitem[{Mompiou et~al.(2015)Mompiou, Legros, Ensslen and
  Kraft}]{Mompiou2015coupling}
\bibinfo{author}{Mompiou, F.}, \bibinfo{author}{Legros, M.},
  \bibinfo{author}{Ensslen, C.}, \bibinfo{author}{Kraft, O.},
  \bibinfo{year}{2015}.
\newblock \bibinfo{title}{In situ tem study of twin boundary migration in
  sub-micron be fibers}.
\newblock \bibinfo{journal}{Acta Materialia} \bibinfo{volume}{96},
  \bibinfo{pages}{57 -- 65}.
\newblock \URLprefix
  \url{http://www.sciencedirect.com/science/article/pii/S1359645415003997},
  \DOIprefix\doi{https://doi.org/10.1016/j.actamat.2015.06.016}.
\bibitem[{Mullins(1956)}]{Mullins1956two}
\bibinfo{author}{Mullins, W.W.}, \bibinfo{year}{1956}.
\newblock \bibinfo{title}{Two-dimensional motion of idealized grain
  boundaries}.
\newblock \bibinfo{journal}{Journal of Applied Physics} \bibinfo{volume}{27},
  \bibinfo{pages}{900--904}.
\newblock \URLprefix \url{https://doi.org/10.1063/1.1722511},
  \DOIprefix\doi{10.1063/1.1722511},
  \href{http://arxiv.org/abs/https://doi.org/10.1063/1.1722511}{{\tt
  arXiv:https://doi.org/10.1063/1.1722511}}.
\bibitem[{Peach and Koehler(1950)}]{PK_force1950}
\bibinfo{author}{Peach, M.}, \bibinfo{author}{Koehler, J.S.},
  \bibinfo{year}{1950}.
\newblock \bibinfo{title}{The forces exerted on dislocations and the stress
  fields produced by them}.
\newblock \bibinfo{journal}{Phys. Rev.} \bibinfo{volume}{80},
  \bibinfo{pages}{436--439}.
\newblock \URLprefix \url{https://link.aps.org/doi/10.1103/PhysRev.80.436},
  \DOIprefix\doi{10.1103/PhysRev.80.436}.
\bibitem[{Rajabzadeh et~al.(2013a)Rajabzadeh, Legros, Combe, Mompiou and
  Molodov}]{Rajabzadeh2013evidence}
\bibinfo{author}{Rajabzadeh, A.}, \bibinfo{author}{Legros, M.},
  \bibinfo{author}{Combe, N.}, \bibinfo{author}{Mompiou, F.},
  \bibinfo{author}{Molodov, D.}, \bibinfo{year}{2013}a.
\newblock \bibinfo{title}{Evidence of grain boundary dislocation step motion
  associated to shear-coupled grain boundary migration}.
\newblock \bibinfo{journal}{Philosophical Magazine} \bibinfo{volume}{93},
  \bibinfo{pages}{1299--1316}.
\newblock \URLprefix \url{https://doi.org/10.1080/14786435.2012.760760},
  \DOIprefix\doi{10.1080/14786435.2012.760760},
  \href{http://arxiv.org/abs/https://doi.org/10.1080/14786435.2012.760760}{{\tt
  arXiv:https://doi.org/10.1080/14786435.2012.760760}}.
\bibitem[{Rajabzadeh et~al.(2013b)Rajabzadeh, Mompiou, Legros and
  Combe}]{rajabzadeh2013elementary}
\bibinfo{author}{Rajabzadeh, A.}, \bibinfo{author}{Mompiou, F.},
  \bibinfo{author}{Legros, M.}, \bibinfo{author}{Combe, N.},
  \bibinfo{year}{2013}b.
\newblock \bibinfo{title}{Elementary mechanisms of shear-coupled grain boundary
  migration}.
\newblock \bibinfo{journal}{Physical Review Letter} \bibinfo{volume}{110},
  \bibinfo{pages}{265507}.
\newblock \URLprefix
  \url{https://link.aps.org/doi/10.1103/PhysRevLett.110.265507},
  \DOIprefix\doi{10.1103/PhysRevLett.110.265507}.
\bibitem[{Read and Shockley(1950)}]{Read1950}
\bibinfo{author}{Read, W.T.}, \bibinfo{author}{Shockley, W.},
  \bibinfo{year}{1950}.
\newblock \bibinfo{title}{Dislocation models of crystal grain boundaries}.
\newblock \bibinfo{journal}{Phys. Rev.} \bibinfo{volume}{78},
  \bibinfo{pages}{275--289}.
\newblock \URLprefix \url{https://link.aps.org/doi/10.1103/PhysRev.78.275},
  \DOIprefix\doi{10.1103/PhysRev.78.275}.
\bibitem[{Riontino et~al.(1979)Riontino, Antonione, Battezzati, Marino and
  Tabasso}]{Riontino1979}
\bibinfo{author}{Riontino, G.}, \bibinfo{author}{Antonione, C.},
  \bibinfo{author}{Battezzati, L.}, \bibinfo{author}{Marino, F.},
  \bibinfo{author}{Tabasso, M.C.}, \bibinfo{year}{1979}.
\newblock \bibinfo{title}{Kinetics of abnormal grain growth in pure iron}.
\newblock \bibinfo{journal}{Journal of Materials Science} \bibinfo{volume}{14},
  \bibinfo{pages}{86--90}.
\newblock \URLprefix \url{https://doi.org/10.1007/BF01028331},
  \DOIprefix\doi{10.1007/BF01028331}.
\bibitem[{Rollett et~al.(1989)Rollett, Srolovitz and
  Anderson}]{rollett1989abnormal}
\bibinfo{author}{Rollett, A.}, \bibinfo{author}{Srolovitz, D.},
  \bibinfo{author}{Anderson, M.}, \bibinfo{year}{1989}.
\newblock \bibinfo{title}{Simulation and theory of abnormal grain
  growth--anisotropic grain boundary energies and mobilities}.
\newblock \bibinfo{journal}{Acta Metallurgica} \bibinfo{volume}{37},
  \bibinfo{pages}{1227 -- 1240}.
\newblock \URLprefix
  \url{http://www.sciencedirect.com/science/article/pii/000161608990117X},
  \DOIprefix\doi{https://doi.org/10.1016/0001-6160(89)90117-X}.
\bibitem[{Rupert et~al.(2009)Rupert, Gianola, Gan and
  Hemker}]{rupert2009science}
\bibinfo{author}{Rupert, T.J.}, \bibinfo{author}{Gianola, D.S.},
  \bibinfo{author}{Gan, Y.}, \bibinfo{author}{Hemker, K.J.},
  \bibinfo{year}{2009}.
\newblock \bibinfo{title}{Experimental observations of stress-driven grain
  boundary migration}.
\newblock \bibinfo{journal}{Science} \bibinfo{volume}{326},
  \bibinfo{pages}{1686--1690}.
\newblock \URLprefix \url{http://science.sciencemag.org/content/326/5960/1686},
  \DOIprefix\doi{10.1126/science.1178226},
  \href{http://arxiv.org/abs/http://science.sciencemag.org/content/326/5960/1686.full.pdf}{{\tt
  arXiv:http://science.sciencemag.org/content/326/5960/1686.full.pdf}}.
\bibitem[{Sansoz and Molinari(2005)}]{Sansoz2005MD}
\bibinfo{author}{Sansoz, F.}, \bibinfo{author}{Molinari, J.},
  \bibinfo{year}{2005}.
\newblock \bibinfo{title}{Mechanical behavior of Σ tilt grain boundaries in
  nanoscale cu and al: A quasicontinuum study}.
\newblock \bibinfo{journal}{Acta Materialia} \bibinfo{volume}{53},
  \bibinfo{pages}{1931 -- 1944}.
\newblock \URLprefix
  \url{http://www.sciencedirect.com/science/article/pii/S1359645405000182},
  \DOIprefix\doi{https://doi.org/10.1016/j.actamat.2005.01.007}.
\bibitem[{Sch{\"{a}}fer and Albe(2012)}]{SCHAFER2012}
\bibinfo{author}{Sch{\"{a}}fer, J.}, \bibinfo{author}{Albe, K.},
  \bibinfo{year}{2012}.
\newblock \bibinfo{title}{Competing deformation mechanisms in nanocrystalline
  metals and alloys: Coupled motion versus grain boundary sliding}.
\newblock \bibinfo{journal}{Acta Materialia} \bibinfo{volume}{60},
  \bibinfo{pages}{6076 -- 6085}.
\newblock \URLprefix
  \url{http://www.sciencedirect.com/science/article/pii/S1359645412004922},
  \DOIprefix\doi{https://doi.org/10.1016/j.actamat.2012.07.044}.
\bibitem[{Shan et~al.(2004)Shan, Stach, Wiezorek, Knapp, Follstaedt and
  Mao}]{Shan2004rotation}
\bibinfo{author}{Shan, Z.}, \bibinfo{author}{Stach, E.A.},
  \bibinfo{author}{Wiezorek, J.M.K.}, \bibinfo{author}{Knapp, J.A.},
  \bibinfo{author}{Follstaedt, D.M.}, \bibinfo{author}{Mao, S.X.},
  \bibinfo{year}{2004}.
\newblock \bibinfo{title}{Grain boundary-mediated plasticity in nanocrystalline
  nickel}.
\newblock \bibinfo{journal}{Science} \bibinfo{volume}{305},
  \bibinfo{pages}{654--657}.
\newblock \URLprefix \url{http://science.sciencemag.org/content/305/5684/654},
  \DOIprefix\doi{10.1126/science.1098741},
  \href{http://arxiv.org/abs/http://science.sciencemag.org/content/305/5684/654.full.pdf}{{\tt
  arXiv:http://science.sciencemag.org/content/305/5684/654.full.pdf}}.
\bibitem[{Shiga and Shinoda(2004)}]{Shiga2004MD}
\bibinfo{author}{Shiga, M.}, \bibinfo{author}{Shinoda, W.},
  \bibinfo{year}{2004}.
\newblock \bibinfo{title}{Stress-assisted grain boundary sliding and migration
  at finite temperature: A molecular dynamics study}.
\newblock \bibinfo{journal}{Physical Review B} \bibinfo{volume}{70},
  \bibinfo{pages}{054102}.
\newblock \URLprefix \url{https://link.aps.org/doi/10.1103/PhysRevB.70.054102},
  \DOIprefix\doi{10.1103/PhysRevB.70.054102}.
\bibitem[{Simpson et~al.(1971)Simpson, Aust and Winegard}]{Simpson1971}
\bibinfo{author}{Simpson, C.J.}, \bibinfo{author}{Aust, K.T.},
  \bibinfo{author}{Winegard, W.C.}, \bibinfo{year}{1971}.
\newblock \bibinfo{title}{The four stages of grain growth}.
\newblock \bibinfo{journal}{Metallurgical Transactions} \bibinfo{volume}{2},
  \bibinfo{pages}{987--991}.
\newblock \URLprefix \url{https://doi.org/10.1007/BF02664229},
  \DOIprefix\doi{10.1007/BF02664229}.
\bibitem[{Srinivasan and Cahn(2002)}]{srinivasan2002rotation}
\bibinfo{author}{Srinivasan, S.}, \bibinfo{author}{Cahn, J.},
  \bibinfo{year}{2002}.
\newblock \bibinfo{title}{Challenging Some Free-Energy Reduction Criteria for
  Grain Growth}. \bibinfo{publisher}{John Wiley \& Sons, Ltd}.
\newblock pp. \bibinfo{pages}{1--14}.
\newblock \URLprefix
  \url{https://onlinelibrary.wiley.com/doi/abs/10.1002/9781118788103.ch1},
  \DOIprefix\doi{10.1002/9781118788103.ch1},
  \href{http://arxiv.org/abs/https://onlinelibrary.wiley.com/doi/pdf/10.1002/9781118788103.ch1}{{\tt
  arXiv:https://onlinelibrary.wiley.com/doi/pdf/10.1002/9781118788103.ch1}}.
\bibitem[{Sutton and Balluffi(1995)}]{sutton1995interfaces}
\bibinfo{author}{Sutton, A.}, \bibinfo{author}{Balluffi, R.},
  \bibinfo{year}{1995}.
\newblock \bibinfo{title}{Interfaces in Crystalline Materials}.
\newblock \bibinfo{publisher}{Oxford: New York: Clarendon Press}.
\newblock \URLprefix \url{https://books.google.com.hk/books?id=VLFqAAAACAAJ}.
\bibitem[{Taylor and Cahn(2007)}]{taylor2007accommodation}
\bibinfo{author}{Taylor, J.E.}, \bibinfo{author}{Cahn, J.},
  \bibinfo{year}{2007}.
\newblock \bibinfo{title}{Shape accommodation of a rotating embedded crystal
  via a new variational formulation}.
\newblock \bibinfo{journal}{Interfaces and Free Boundaries}
  \bibinfo{volume}{9}, \bibinfo{pages}{493--512}.
\newblock \DOIprefix\doi{10.4171/IFB/174}.
\bibitem[{Thomas et~al.(2017)Thomas, Chen, Han, Purohit and
  Srolovitz}]{thomas2017reconciling}
\bibinfo{author}{Thomas, S.L.}, \bibinfo{author}{Chen, K.},
  \bibinfo{author}{Han, J.}, \bibinfo{author}{Purohit, P.K.},
  \bibinfo{author}{Srolovitz, D.J.}, \bibinfo{year}{2017}.
\newblock \bibinfo{title}{Reconciling grain growth and shear-coupled grain
  boundary migration}.
\newblock \bibinfo{journal}{Nature communications} \bibinfo{volume}{8},
  \bibinfo{pages}{1764}.
\bibitem[{Thomas et~al.(2019)Thomas, Wei, Han, Xiang and
  Srolovitz}]{thomas2019TJ}
\bibinfo{author}{Thomas, S.L.}, \bibinfo{author}{Wei, C.},
  \bibinfo{author}{Han, J.}, \bibinfo{author}{Xiang, Y.},
  \bibinfo{author}{Srolovitz, D.J.}, \bibinfo{year}{2019}.
\newblock \bibinfo{title}{Disconnection description of triple-junction motion}.
\newblock \bibinfo{journal}{Proceedings of the National Academy of Sciences}
  \bibinfo{volume}{116}, \bibinfo{pages}{8756--8765}.
\newblock \URLprefix \url{https://www.pnas.org/content/116/18/8756},
  \DOIprefix\doi{10.1073/pnas.1820789116},
  \href{http://arxiv.org/abs/https://www.pnas.org/content/116/18/8756.full.pdf}{{\tt
  arXiv:https://www.pnas.org/content/116/18/8756.full.pdf}}.
\bibitem[{Trautt et~al.(2012)Trautt, Adland, Karma and Mishin}]{Trautt2012MD}
\bibinfo{author}{Trautt, Z.}, \bibinfo{author}{Adland, A.},
  \bibinfo{author}{Karma, A.}, \bibinfo{author}{Mishin, Y.},
  \bibinfo{year}{2012}.
\newblock \bibinfo{title}{Coupled motion of asymmetrical tilt grain boundaries:
  Molecular dynamics and phase field crystal simulations}.
\newblock \bibinfo{journal}{Acta Materialia} \bibinfo{volume}{60},
  \bibinfo{pages}{6528 -- 6546}.
\newblock \URLprefix
  \url{http://www.sciencedirect.com/science/article/pii/S1359645412005472},
  \DOIprefix\doi{https://doi.org/10.1016/j.actamat.2012.08.018}.
\bibitem[{Upmanyu et~al.(2002)Upmanyu, Hassold, Kazaryan, Holm, Wang, Patton
  and Srolovitz}]{upmanyu2002simulation}
\bibinfo{author}{Upmanyu, M.}, \bibinfo{author}{Hassold, G.},
  \bibinfo{author}{Kazaryan, A.}, \bibinfo{author}{Holm, E.},
  \bibinfo{author}{Wang, Y.}, \bibinfo{author}{Patton, B.},
  \bibinfo{author}{Srolovitz, D.}, \bibinfo{year}{2002}.
\newblock \bibinfo{title}{Boundary mobility and energy anisotropy effects on
  microstructural evolution during grain growth}.
\newblock \bibinfo{journal}{Interface Science} \bibinfo{volume}{10},
  \bibinfo{pages}{201--216}.
\newblock \URLprefix \url{https://doi.org/10.1023/A:1015832431826},
  \DOIprefix\doi{10.1023/A:1015832431826}.
\bibitem[{Upmanyu et~al.(1998)Upmanyu, Smith and Srolovitz}]{upmanyu1998}
\bibinfo{author}{Upmanyu, M.}, \bibinfo{author}{Smith, R.},
  \bibinfo{author}{Srolovitz, D.}, \bibinfo{year}{1998}.
\newblock \bibinfo{title}{Atomistic simulation of curvature driven grain
  boundary migration}.
\newblock \bibinfo{journal}{Interface Science} \bibinfo{volume}{6},
  \bibinfo{pages}{41--58}.
\newblock \URLprefix \url{https://doi.org/10.1023/A:1008608418845},
  \DOIprefix\doi{10.1023/A:1008608418845}.
\bibitem[{Upmanyu et~al.(2006)Upmanyu, Srolovitz, Lobkovsky, Warren and
  Carter}]{upmanyu2006simulation}
\bibinfo{author}{Upmanyu, M.}, \bibinfo{author}{Srolovitz, D.},
  \bibinfo{author}{Lobkovsky, A.}, \bibinfo{author}{Warren, J.},
  \bibinfo{author}{Carter, W.}, \bibinfo{year}{2006}.
\newblock \bibinfo{title}{Simultaneous grain boundary migration and grain
  rotation}.
\newblock \bibinfo{journal}{Acta Materialia} \bibinfo{volume}{54},
  \bibinfo{pages}{1707 -- 1719}.
\newblock \URLprefix
  \url{http://www.sciencedirect.com/science/article/pii/S135964540500707X},
  \DOIprefix\doi{https://doi.org/10.1016/j.actamat.2005.11.036}.
\bibitem[{Van~Swygenhoven et~al.(2002)Van~Swygenhoven, Derlet and
  Hasnaoui}]{VanSwygenhoven2002dislocationemission}
\bibinfo{author}{Van~Swygenhoven, H.}, \bibinfo{author}{Derlet, P.M.},
  \bibinfo{author}{Hasnaoui, A.}, \bibinfo{year}{2002}.
\newblock \bibinfo{title}{Atomic mechanism for dislocation emission from
  nanosized grain boundaries}.
\newblock \bibinfo{journal}{Physical Review B} \bibinfo{volume}{66},
  \bibinfo{pages}{024101}.
\newblock \URLprefix \url{https://link.aps.org/doi/10.1103/PhysRevB.66.024101},
  \DOIprefix\doi{10.1103/PhysRevB.66.024101}.
\bibitem[{Von~Neumann(1952)}]{von1952metal}
\bibinfo{author}{Von~Neumann, J.}, \bibinfo{year}{1952}.
\newblock \bibinfo{title}{Metal interfaces}.
\newblock \bibinfo{journal}{American Society for Metals, Cleveland}
  \bibinfo{volume}{108}.
\bibitem[{Winning et~al.(2001)Winning, Gottstein and
  Shvindlerman}]{winning2001stressinduced}
\bibinfo{author}{Winning, M.}, \bibinfo{author}{Gottstein, G.},
  \bibinfo{author}{Shvindlerman, L.}, \bibinfo{year}{2001}.
\newblock \bibinfo{title}{Stress induced grain boundary motion}.
\newblock \bibinfo{journal}{Acta Materialia} \bibinfo{volume}{49},
  \bibinfo{pages}{211 -- 219}.
\newblock \URLprefix
  \url{http://www.sciencedirect.com/science/article/pii/S1359645400003219},
  \DOIprefix\doi{https://doi.org/10.1016/S1359-6454(00)00321-9}.
\bibitem[{Winning et~al.(2002)Winning, Gottstein and
  Shvindlerman}]{winning2002mechanism}
\bibinfo{author}{Winning, M.}, \bibinfo{author}{Gottstein, G.},
  \bibinfo{author}{Shvindlerman, L.}, \bibinfo{year}{2002}.
\newblock \bibinfo{title}{On the mechanisms of grain boundary migration}.
\newblock \bibinfo{journal}{Acta Materialia} \bibinfo{volume}{50},
  \bibinfo{pages}{353 -- 363}.
\newblock \URLprefix
  \url{http://www.sciencedirect.com/science/article/pii/S1359645401003433},
  \DOIprefix\doi{https://doi.org/10.1016/S1359-6454(01)00343-3}.
\bibitem[{Yang et~al.(2015)Yang, Mao, Li and Liu}]{yang2015}
\bibinfo{author}{Yang, J.Z.}, \bibinfo{author}{Mao, C.}, \bibinfo{author}{Li,
  X.}, \bibinfo{author}{Liu, C.}, \bibinfo{year}{2015}.
\newblock \bibinfo{title}{On the cauchy\-born approximation at finite
  temperature}.
\newblock \bibinfo{journal}{Computational Materials Science}
  \bibinfo{volume}{99}, \bibinfo{pages}{21 -- 28}.
\newblock \URLprefix
  \url{http://www.sciencedirect.com/science/article/pii/S0927025614007988},
  \DOIprefix\doi{https://doi.org/10.1016/j.commatsci.2014.11.030}.
\bibitem[{Zhang et~al.(2005)Zhang, Upmanyu and Srolovitz}]{zhang2005curvature}
\bibinfo{author}{Zhang, H.}, \bibinfo{author}{Upmanyu, M.},
  \bibinfo{author}{Srolovitz, D.}, \bibinfo{year}{2005}.
\newblock \bibinfo{title}{Curvature driven grain boundary migration in
  aluminum: molecular dynamics simulations}.
\newblock \bibinfo{journal}{Acta Materialia} \bibinfo{volume}{53},
  \bibinfo{pages}{79 -- 86}.
\newblock \URLprefix
  \url{http://www.sciencedirect.com/science/article/pii/S1359645404005452},
  \DOIprefix\doi{https://doi.org/10.1016/j.actamat.2004.09.004}.
\bibitem[{Zhang et~al.(2017)Zhang, Han, Xiang and Srolovitz}]{zhang2017prl}
\bibinfo{author}{Zhang, L.}, \bibinfo{author}{Han, J.}, \bibinfo{author}{Xiang,
  Y.}, \bibinfo{author}{Srolovitz, D.J.}, \bibinfo{year}{2017}.
\newblock \bibinfo{title}{Equation of motion for a grain boundary}.
\newblock \bibinfo{journal}{Physical review letters} \bibinfo{volume}{119},
  \bibinfo{pages}{246101}.
\bibitem[{Zhang and Xiang(2018)}]{zhang2018lowangle}
\bibinfo{author}{Zhang, L.}, \bibinfo{author}{Xiang, Y.}, \bibinfo{year}{2018}.
\newblock \bibinfo{title}{Motion of grain boundaries incorporating dislocation
  structure}.
\newblock \bibinfo{journal}{Journal of the Mechanics and Physics of Solids}
  \bibinfo{volume}{117}, \bibinfo{pages}{157--178}.

\end{thebibliography}







\end{document}